\documentclass[prb,reprint,superscriptaddress]{revtex4-1}

  \usepackage{amsmath}
  \usepackage{makeidx}
  \usepackage{amsfonts}
  \usepackage[ansinew]{inputenc}
  \usepackage[usenames,dvipsnames]{pstricks}
  \usepackage{subfigure}
  \usepackage{multirow}
  \usepackage{array}
  \usepackage{epsfig}
  \usepackage{tabularx}
  \usepackage{pst-grad} 
  \usepackage{pst-plot} 
  \usepackage[colorlinks,hyperindex]{hyperref}
  \usepackage{siunitx}
  \hypersetup
  {
    colorlinks,%
    citecolor=black,%
    linkcolor=black,%
    urlcolor=black,%
  }

  \newcommand{\abs}[1]{\left| #1 \right|}
  \newcommand{\Ams}{\mathrm{\AA}}
  \renewcommand{\vec}[1]{\boldsymbol{\mathbf{#1}}}
  \renewcommand{\exp}[1]{\mathrm{e}^{#1}}
  \newcommand{\kk}{\mathbf{k}}
  \newcommand{\qq}{\mathbf{q}}
  \newcommand{\KK}{\mathbf{K}}
  \newcommand{\GG}{\mathbf{G}}
  
  \newcommand{\rr}{\mathbf{r}}
  \newcommand{\xxi}{\boldsymbol{\xi}}
  
  \newcommand{\ket}[1]{\left| #1 \right>}

  \newcommand{\braoketf}[3]{\big<  #1 \big| #2 \big| #3 \big>}

\makeindex

\begin{document}

\title{Localized interlayer complexes in heterobilayer transition metal dichalcogenides}

\author{M.\ Danovich}
\affiliation{National Graphene Institute, University of Manchester,
  Booth St E, Manchester M13 9PL, United Kingdom}

\author{D.\ A.\ Ruiz-Tijerina}
\affiliation{National Graphene Institute, University of Manchester,
  Booth St E, Manchester M13 9PL, United Kingdom}

\author{R.\ J.\ Hunt}
\affiliation{Department of Physics, Lancaster University, Lancaster
  LA1 4YB, United Kingdom}

\author{M.\ Szyniszewski}
\affiliation{Department of Physics, Lancaster University, Lancaster
  LA1 4YB, United Kingdom}

\author{N.\ D.\ Drummond}
\affiliation{Department of Physics, Lancaster University, Lancaster
  LA1 4YB, United Kingdom}

\author{V.\ I.\ Fal'ko}
\affiliation{National Graphene Institute, University of Manchester,
  Booth St E, Manchester M13 9PL, United Kingdom}


\date{\today}

\begin{abstract}
We present theoretical results for the radiative rates and
doping-dependent photoluminescence spectrum of interlayer excitonic complexes
localized by donor impurities in MoSe$_2$/WSe$_2$ twisted
heterobilayers, supported by quantum Monte Carlo calculations of
binding energies and wave-function overlap integrals.  For closely
aligned layers, radiative decay is made possible by the momentum
spread of the localized complexes' wave functions, resulting in radiative rates
of a few
$\mu$s$^{-1}$. For strongly misaligned layers, the
short-range interaction between the carriers and impurity provides a
finite radiative rate with a strong asymptotic twist angle dependence
 $\propto \theta^{-8}$. Finally, phonon-assisted recombination is
considered, with emission of optical phonons in both layers resulting in
additional, weaker emission lines, red shifted by the phonon energy.
\end{abstract}


\maketitle

\section{Introduction \label{sec:intro}}

Recent advances in the study of two-dimensional (2D) materials have
allowed the realization of van der Waals (vdW) heterostructures
consisting of vertically stacked 2D layers, resulting in unique
properties and potential novel device applications
\cite{geim_hetero,novoselov_hetero, zhang_hetero_opto, photovoltaic,pn_hetero}. The layers
forming these heterostructures are only weakly bound by vdW forces,
and largely retain their individual characteristic properties. Yet,
the weak interlayer coupling allows the different properties of
various 2D materials to be combined.

One such family of vdW heterostructures are heterobilayers of
2D transition metal dichalcogenides (TMDs), which have attracted much interest due to their unique optical properties, dominated
by strongly bound excitonic complexes
\cite{marcin_bindingenergies_prb_2017, high_exciton_binding_energy}
and spin- and valley-dependent optical selection rules \cite{Cao2012,
  Xu2014}.  The most commonly studied heterobilayers are of the form
MoX$_2$/WX$_2$, with X${}={}$S or Se, due to their type-II (staggered)
band alignment, in which the lowest conduction-band (CB) edge and the highest valence band (VB) edge are spatially confined to different layers
 \cite{band_alignment1, band_alignment2}. In this configuration,
 electrostatic interactions between electrons and holes across the heterostructure result in the formation of interlayer excitonic
 complexes, whose constituent carriers are spatially separated in
 the out-of-plane direction. Optical signatures of these interlayer
 complexes have been reported in photoluminescence (PL) experiments
 \cite{rivera_interlayer, nagler2016_interlayer_excitons,alexeev_nanolett_2017}, where new
 PL peaks are observed in the spectra of bilayer regions. These
 signatures appear at energies below the monolayer photoemission
 lines, due to the smaller interlayer band gap in the staggered band
 configuration.
\begin{figure}[t!]
  \centering
  \includegraphics[width=1\columnwidth]{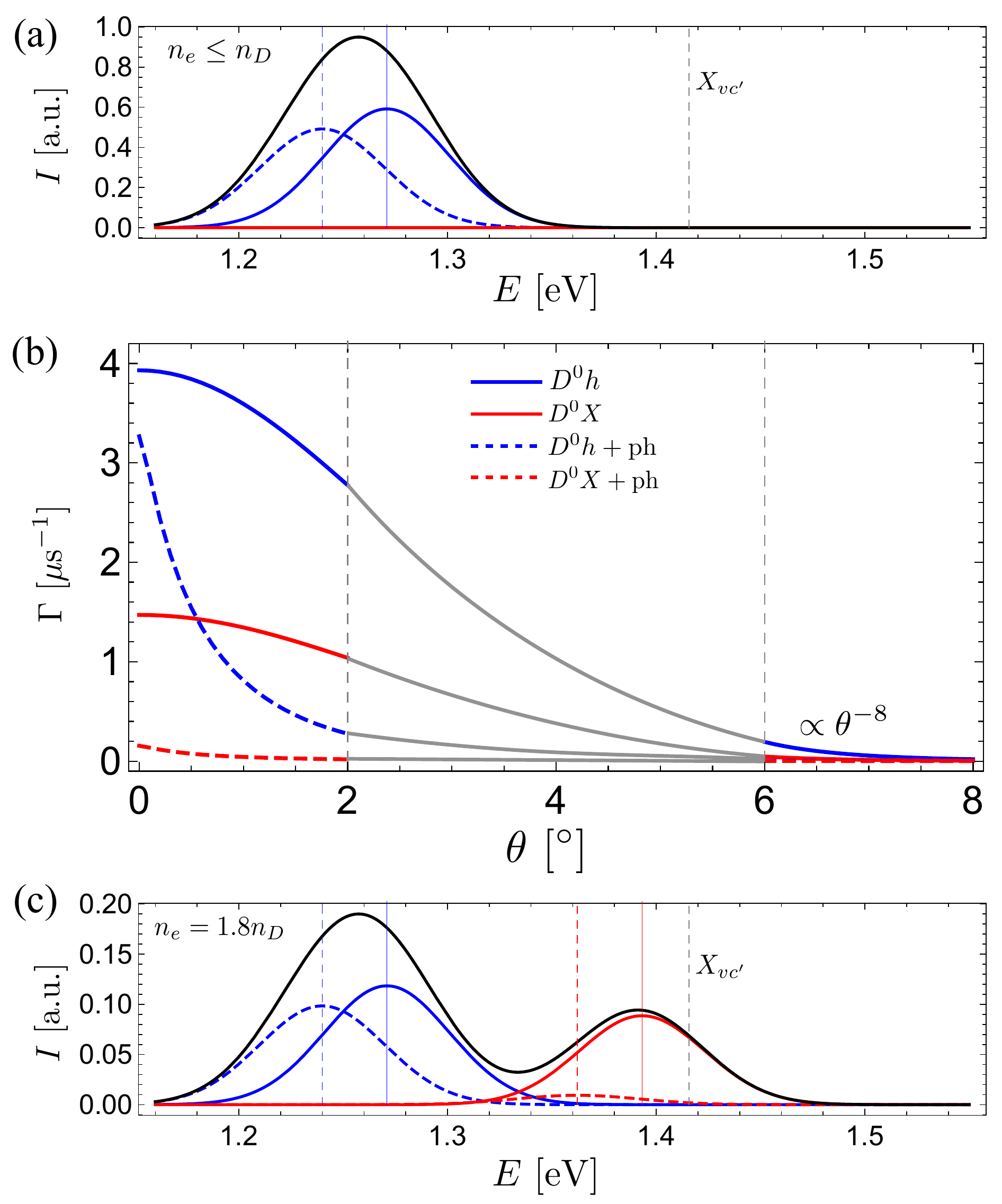}
  \caption{ (a) Simulated PL spectrum of donor-bound interlayer
          complexes in an aligned ($\theta=0$) ${\rm MoSe_2/WSe_2}$
          bilayer encapsulated in hexagonal boron nitride, for an electron
          density of $n_{\rm e}=0.9n_{\rm D}$, with $n_{\rm D}$ being the donor
          density. Dashed lines indicate PL from phonon-assisted
          recombination. Solid lines are taken to have Gaussian shape with width $2\sigma=60$ meV, and the interlayer gap is $\tilde{E}_g=1.5$ eV\@. The vertical gray dashed lines in (a) and (c) indicate the position of the free interlayer exciton $X_{vc'}$.
          (b) Radiative rates of the D$^0_{c'}$h$_{v}$ (per hole) (solid
          blue) and D$^0_{c'}$X$_{vc'}$ (solid red) complexes, and their
          phonon-assisted replicas (dashed), in the large and small
          twist angle ($\theta$) limits. The rates have a strong
          angular dependence, with asymptotic behavior $\sim
          \theta^{-8}$ for radiative decay driven by short-range
          interactions, and $\sim \theta^{-4}$ for phonon-assisted
          processes. The gray lines for intermediate twist angles $\theta=2$--$6^{\circ}$ have been interpolated by hand.
          (c) Simulated PL spectrum in the limit of heavy
          $n$-doping, showing the appearance of the donor-bound trion
          (D$^0_{c'}$X$_{vc'}$) line when $n_{\rm e}>n_{\rm D}$. Parameters:
          $n_{\rm h} = 10^{11}$ cm$^{-2}$ and $n_{\rm D}=10^{13}$ cm$^{-2}$.}
  \label{fig:figure1}
\end{figure}

Photoemission by free interlayer excitons is limited by the relative
interlayer angle $\theta$ and the incommensurability of the two TMD
lattices $\delta$, resulting in a momentum-space mismatch
$\Delta K\approx K\sqrt{\delta^2+\theta^2}$ between the conduction-
and valence-band edges, as shown in Fig.\ \ref{fig:alignment}(b).
Radiative recombination becomes effectively indirect, and thus
suppressed by energy and momentum conservation
\cite{wangyao_anom}. These constraints are relaxed when interlayer
excitons and larger excitonic complexes localize about charged
defects, such as donor ions, which are commonly observed as dopants in
real samples. Formation of these complexes is favored by the long
interlayer exciton lifetimes resulting from the spatial
separation of their carriers, which allow for their localization by the deep potential
wells provided by the ions. The spread in momentum space
of these localized complexes opens the possibility for a finite radiative matrix
element $M\propto \int d^2 r \, e^{i \Delta {\bf K}\cdot {\bf r}}
\Psi(\rr)$, where $\Psi(\rr)$ is the envelope wave function of the
complex.

In this paper, we provide a theory for the radiative recombination of
localized interlayer complexes in TMD heterostructures of the form
MoX${}_2$/WX${}_2$, where the carriers are bound to a donor ion in the
MoX$_2$ layer.  Focusing specifically on MoSe${}_2$/WSe${}_2$ encapsulated in hexagonal boron nitride (hBN), we use
variational and diffusion quantum Monte Carlo (VMC and DMC)
simulations \cite{Ceperley_1980,Foulkes_2001} to evaluate the binding
energies and wave-function overlap integrals of complexes involving one or
two holes in the WSe$_2$ layer and up to four electrons in the
MoSe$_2$ layer, accounting for bilayer and encapsulation screening effects.
We discuss the energetics and stability of these complexes based on their
binding energies, and the robustness of our results against uncertainty in model parameters,
such as the carrier effective masses and screening lengths.

Motivated by the binding energies obtained from our quantum Monte
Carlo (QMC) calculations and PL experiments \cite{vialla_private},
we study the radiative recombination of the two simplest complexes
consisting of MoX${}_2$ electrons and a single WX${}_2$ hole bound to an
impurity center: a donor ion and an exciton (D$^0_{c'}$h$_{v}$), and a
donor-bound trion (D$^0_{c'}$X$_{vc'}$). We predict the qualitative PL spectrum from these complexes for closely aligned
TMD heterobilayers, and estimate the asymptotic behavior of their PL signals in the
regime of strong misalignment based on general kinematics
and perturbation theory. Our results indicate a rapid decay of the PL signals from
the most relevant donor-bound interlayer complexes with the interlayer twist angle
($\theta$), resulting from the asymptotic behavior $\Gamma \sim \theta^{-8}$ of the
radiative rates at strong misalignment. As a consequence, we expect that optical signatures from
these complexes can be detected only in closely aligned crystals.
Our results provide a new perspective for the interpretation of recently reported
luminescence spectra of closely-aligned TMD heterobilayers, where the interlayer
portion of the spectrum has been attributed to delocalized interlayer exciton states \cite{rivera_interlayer}.

The remainder of this paper is organized as follows. In
Sec.\ \ref{sec:model} we discuss the model Hamiltonian for the TMD
heterobilayer, describe our approach to calculating its optical
properties, and present our DMC results for the binding energies of
the main interlayer impurity-bound complexes. In
Sec.\ \ref{sec:complexes}, we address the PL signatures of these
complexes, assuming good alignment between the TMD monolayers in the
heterostructure, and estimate the asymptotic behavior of their radiative decay
with twist angle in Sec.\ \ref{sec:asymptotics}. We consider the
effects of electron-phonon interactions in
Sec.\ \ref{sec:e_ph_effects}, and we find that longitudinal optical
phonon modes can introduce red shifted replicas to the main PL
lines. Finally, we estimate the evolution of the PL spectrum of the
two main donor-bound interlayer complexes with doping in
Sec.\ \ref{sec:intens}. Our conclusions are summarized in
Fig.\ \ref{fig:figure1}, and discussed in Sec.\ \ref{sec:conclusions}.

\section{Model}\label{sec:model}
\subsection{Electrostatic interactions in a bilayer system}\label{sec:model_bilayer_int}
The reduced dimensionality of a
monolayer TMD leads to modified electrostatic interactions between its
charge carriers below a characteristic length scale $r_*=2\pi
\kappa/\epsilon$ (in Gaussian units), determined by the monolayer's
in-plane dielectric susceptibility $\kappa$, and the (average)
dielectric constant $\epsilon$ of its environment \cite{keldysh,ganchev}. In a
TMD heterobilayer, further screening effects must be
considered. The resulting interactions between same-layer carriers
$\mathcal{V}$ in one layer and $\mathcal{V}'$ in the other, and the
interlayer interaction $\mathcal{W}$, have Fourier components
(Appendix \ref{sec:multilayer_keldysh})
\begin{subequations}
\begin{equation}\label{eq:exact_int_11}
  \mathcal{V}({\bf q}) = \frac{2\pi\left(1+r_{*}'q - r_{*}'q\exp{-2qd}\right)}
  {\epsilon q\left[(1+r_{*}q)(1+r_{*}'q) - r_{*}r_{*}'q^2
  \exp{-2qd}\right]},
  \end{equation}
  \begin{equation}\label{eq:exact_int_22}
 \mathcal{V}'({\bf q}) = \frac{2\pi\left(1+r_{*}q - r_{*}q\exp{-2qd}\right)}
  {\epsilon q\left[(1+r_{*}q)(1+r_{*}'q) - r_{*}r_{*}'q^2
  \exp{-2qd}\right]},
  \end{equation}
  \begin{equation}\label{eq:exact_int_12}
  \mathcal{W}({\bf q}) = \frac{2\pi\,\exp{-qd}}
  {\epsilon q\left[(1+r_{*}q)(1+r_{*}'q) - r_{*}r_{*}'q^2
  \exp{-2qd}\right]},
\end{equation}
\end{subequations}
where $\qq$ is the wave vector, $d$ is the interlayer
distance, and $r_{*}$ and $r_*'$ are the corresponding monolayer screening lengths.

Previous works on monolayer TMDs have focused on interactions of the
Keldysh form \cite{keldysh} to study their excitonic spectra and
optical properties
\cite{berkelbach_prb_2013,ganchev,vanderdonk_prb_2017,marcin_bindingenergies_prb_2017,mostaani_excitonic_prb_2017}. For
bilayers, this potential form is obtained from
Eqs.\ (\ref{eq:exact_int_11})--(\ref{eq:exact_int_12}) in the
long-range limit ($q\ll 1/r_{*},1/r_*'$) as
\begin{subequations}
\begin{equation}\label{eq:keldysh_form_nn}
  \mathcal{V}_{<}(\qq)=\mathcal{V}_{<}'(\qq)=\frac{2\pi
        }{\epsilon q\left[1 + (r_{*}+r_{*}')q \right]},
\end{equation}
\begin{equation}\label{eq:keldysh_form_12}
  \mathcal{W}_{<}(\qq)=\frac{2\pi }{\epsilon q\left[1 +
            (r_{*}+r_{*}'+d)q \right]}.
\end{equation}
\end{subequations}
By contrast, in the short-range limit ($q\gg 1/r_{*},1/r_{*}'$) we
obtain for the intralayer interactions
\begin{equation}\label{eq:short_range_intra}
  \mathcal{V}_{>}(\qq)=\frac{2\pi }{\epsilon\, r_{*}q^2}, \quad
        \mathcal{V}_{>}'(\qq)=\frac{2\pi }{\epsilon\, r_{*}'q^2},
\end{equation}
revealing the absence of screening from the opposite layer in this
regime. More strikingly, the short-range interlayer potential vanishes
exponentially as $\mathcal{W}_{>}(\qq)=2\pi\,\exp{-qd}/( \epsilon\,
r_{*}r_{*}'q^3)$. Neither of these features is captured by
extrapolation of Eqs.\ (\ref{eq:keldysh_form_nn}) and
(\ref{eq:keldysh_form_12}) to large wave numbers.

\begin{figure}[t!]
  \centering
  \includegraphics[width=\columnwidth]{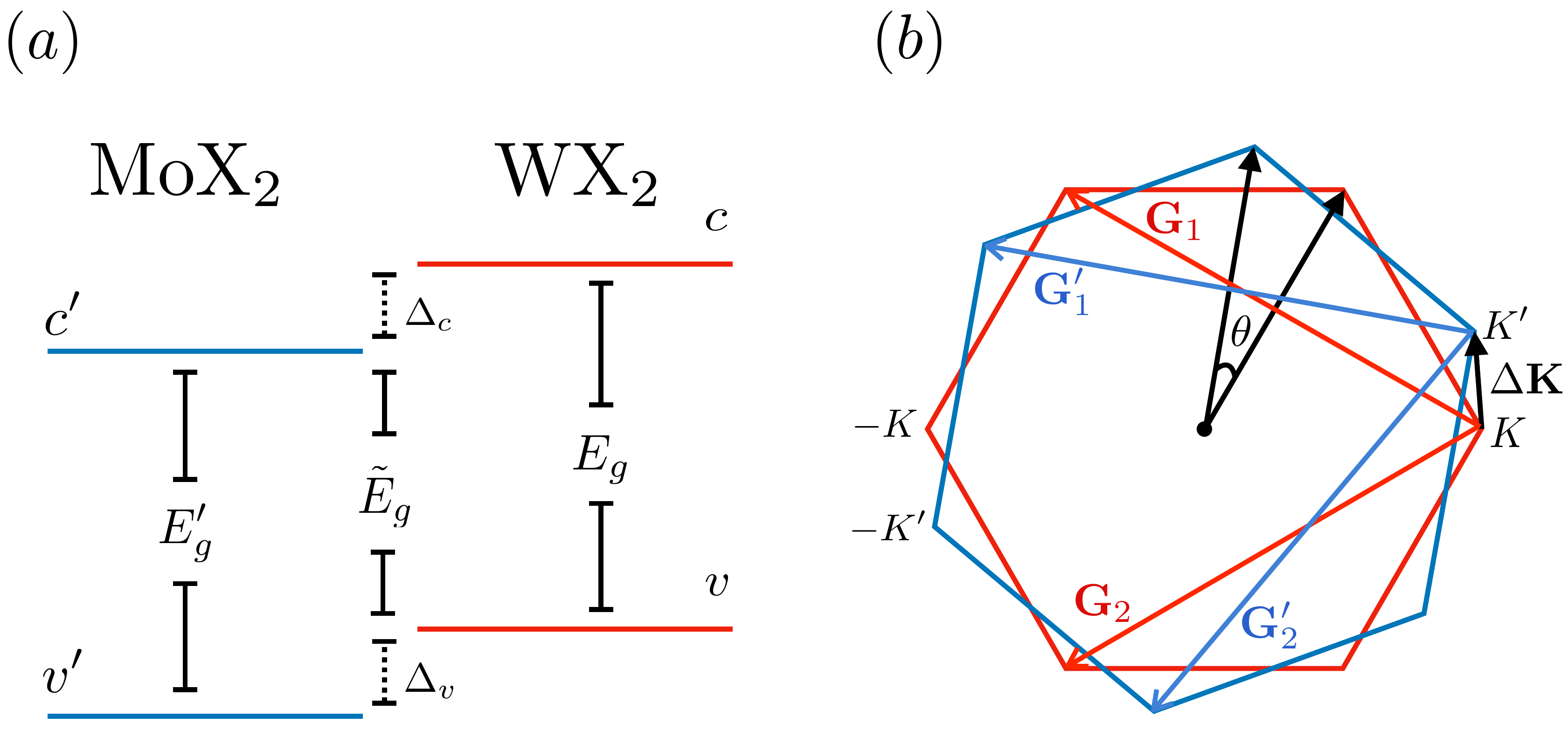}
  \caption{ (a) Schematic of type-II band alignment in a TMD
          heterobilayer. The CB and VB of the two layers are shifted
          relative to each other by energies $\Delta_c$ and
          $\Delta_v$, respectively, giving an interlayer gap of
          $\tilde{E}_{\rm g}$. (b) The Brillouin zones (BZs) of the
          misaligned TMD monolayers, with $\GG_n$ and $\GG_n'$ their
          main reciprocal lattice vectors.
          Their $K$ valleys are separated
          by a momentum vector $\Delta {\bf K}$, due to the nonzero
          misalignment angle $\theta$ and to the difference in lattice
          constants.  }
  \label{fig:alignment}
\end{figure}

\subsection{Photon emission by donor-bound complexes}\label{sec:model_donor_bound}

As in the monolayer case
\cite{ramasub_prb_2012,mak_natmat_2012,ross_natcomm_2013,palacios_nanolett_2014},
optical properties of the heterobilayer are determined by excitonic
complexes formed by excess electrons and holes in the
sample. Staggered (type-II) band alignment, in which the main electron
and hole bands belong to opposite layers, is typical of TMD
heterostructures \cite{band_alignment1}. This is shown schematically
in Fig.\ \ref{fig:alignment}(a) for a MoX${}_2$/WX${}_2$ structure,
where X${}={}$S or Se represents a chalcogen; the main electron and hole bands
are labeled $c'$ and $v$, respectively, and the primed (unprimed) band
labels correspond to the MoX${}_2$ (WX${}_2$) layer. Given the reduced band gap $\tilde{E}_g$ [Fig.\ \ref{fig:alignment}(a)], the lowest-energy exciton states are spread across the heterostructure,
formed by $c'$-band electrons and $v$-band holes bound by
the interaction $\mathcal{W}(\qq)$
\cite{rivera_interlayer,yifei_nanolett_2015,nagler2016_interlayer_excitons}.

The optical activity of interlayer excitons in TMD bilayers is
strongly constrained by the interlayer alignment. As shown in
Fig.\ \ref{fig:alignment}(b), the relative twist angle and lattice
incommensurability between the two layers produces a mismatch between
their Brillouin zones (BZs). Thus, bright interlayer excitons in
MoX${}_2$/WX${}_2$ structures, consisting of same-valley $c'$-band
electrons and $v$-band holes, have a finite center-of-mass momentum
$\Delta\KK = \KK' - \KK$. Due to energy and momentum conservation, photon emission by interlayer excitons is only allowed when \cite{rana_prb_2016}
$\Delta K \approx 0$.

The above restrictions are relaxed when excitons and other excitonic
complexes are bound to impurity centers in the sample, such as charged
defects and donor ions. These complexes are localized within some
characteristic length $a_0^*$, the Bohr radius of the complex, such that their momentum-space wave
functions are finite up to momenta of order $1/a_0^*$. As a result, the recombination rates of
impurity-bound interlayer complexes are determined by the
large-momentum tail of their wave function, and thus by the
short-range interaction [Eq.\ (\ref{eq:short_range_intra})].

The Hamiltonian for the heterobilayer in the free-carrier
basis is
\begin{equation}
\hat{H} = \hat{H}_0 + \hat{H}_{\rm t} + \hat{U}_{\text{intra}} + \hat{U}_{\text{inter}},
\label{eq:H}
\end{equation}
where the zeroth-order Hamiltonian $\hat{H}_0$, describing the CB and VB
electrons of the two individual layers, is given in second
quantization as
\begin{equation}
\hat{H}_0 = \sum_{\alpha}\sum_{\kk,\tau,\sigma} E_{\alpha}(\kk)
c^{\dagger}_{\alpha,\tau,\sigma}(\kk) c_{\alpha,\tau,\sigma}(\kk).
\end{equation}
$c^{\dagger}_{\alpha,\tau,\sigma}(\kk)$ creates an electron of spin
projection $\sigma=\uparrow,\,\downarrow$ and momentum $\kk$ relative
to the $\tau\KK$ valley ($\tau=\pm$) of band $\alpha = c',v',c,v$.
The band dispersions are
\begin{subequations}
\begin{equation}\label{eq:Evp}
  E_{v'}(\kk) = -\Delta_v-\frac{\hbar^2k^2}{2m_v'},
\end{equation}
\begin{equation}\label{eq:Ev}
  E_v(\kk) = -\frac{\hbar^2k^2}{2m_v},
\end{equation}
\begin{equation}\label{eq:Ecp}
  E_{c'}(\kk) = \tilde{E}_g+\frac{\hbar^2k^2}{2m_{c'}},
\end{equation}
\begin{equation}\label{eq:Ec}
  E_{c}(\kk) = \tilde{E}_g+\Delta_c+\frac{\hbar^2k^2}{2m_{c}},
\end{equation}
\end{subequations}
where $\Delta_c$ ($\Delta_v$)
is the spacing between the electron (hole) band edges
[Fig.\ \ref{fig:alignment}(a)].

The tunneling Hamiltonian, describing electron hopping between the
layers, is given by\cite{koshino_incommensurate,wangyao_coupling}
\begin{equation}\label{eq:hopping}
\begin{split}
&\hat{H}_{\rm t}=\sum_{\tau,\sigma}\sum_{\GG,\GG'}\sum_{\kk,\kk'}
  \delta_{\tau \KK+\kk+{\bf G},\tau' \KK'+\kk'+{\bf
      G'}}\exp{-i\GG_0\cdot\rr_0}\\ &\times\left[ t_{cc}(\kk+\tau\KK+\GG)
    c^{\dagger}_{c\tau\sigma}(\kk)c_{c'\tau'\sigma}(\kk') \right.\\
    & \qquad+ \left.t_{vv}(\kk+\tau\KK+\GG)
    c^{\dagger}_{v\tau\sigma}(\kk)c_{v'\tau'\sigma}(\kk')\right] +
  \text{H.c.},
\end{split}
\end{equation}
where $t_{cc}(\kk)$ and $t_{vv}(\kk)$ represent interlayer hopping
strengths between the CBs and VBs; ${\bf \GG}$ and ${\bf \GG'}$
correspond to the reciprocal lattice vectors of the hole and electron
layers; and the Kronecker delta enforces momentum conservation in the
tunnelling process. $\rr_0$
  is a vector within the unit cell representing the in-plane shift
  between the metal atoms of the two TMD monolayers, such that a general
  stacking configuration is parameterized by $\rr_0$ and $\theta$. We
focus on configurations with close angular alignment
  but general $\rr_0$; this is a type of pseudo ``AA'' stacking better suited
  to describe experimental situations. Correspondingly, we use the
\emph{ab initio} hopping terms reported in
Ref.\ [\onlinecite{wangyao_coupling}] for AA stacked
  ($\rr_0=0,\,\theta=0$) MoS${}_2$, for estimation purposes. These
values are small (a few meV) compared to all other scales in the
problem, reflecting the vdW and electrical quadrupole nature of the
interlayer interactions. As a result, $\hat{H}_{\rm t}$ can be treated
within perturbation theory. Furthermore, since $t_{\alpha\alpha}(\kk)$
decays rapidly with $k$, we truncate the sums
over $\GG$ and $\GG'$ to the two main Bragg vectors\cite{wangyao_coupling}
  [Fig.\ \ref{fig:alignment}(b)], and set $t_{cc}(\tau\KK) \approx
t_{cc} = 2.5\,\mathrm{meV}$ and $t_{vv}(\tau\KK) \approx t_{vv} =
16\,\mathrm{meV}$.

Finally, the direct electrostatic interactions between carriers, and
between carriers and a positive donor ion of effective charge
$Z_{\text{donor}}$, are given by
\begin{widetext}
\begin{subequations}
\begin{equation}\label{eq:Uintra}
\begin{split}
\hat{U}_{\text{intra}} = \frac{e^2}{S}\sum_{\substack{\tau_1,\tau_2\\\sigma_1,\sigma_2}}\sum_{\kk_1,\kk_2,\xxi}\Bigg[&\sum_{\alpha,\beta=v,c}\frac{\mathcal{V}(\xxi)}{(1+\delta_{\alpha,\beta})}c^{\dagger}_{\alpha,\tau_1,\sigma_1}(\kk_1+\xxi)
c^{\dagger}_{\beta,\tau_2,\sigma_2}(\kk_2-\xxi)
 c_{\beta,\tau_2,\sigma_2}(\kk_2) c_{\alpha,\tau_1,\sigma_1}(\kk_1) \\
 + &\sum_{\alpha,\beta=v',c'}\frac{\mathcal{V}'(\xxi)}{(1+\delta_{\alpha,\beta})}c^{\dagger}_{\alpha,\tau_1,\sigma_1}(\kk_1+\xxi)
c^{\dagger}_{\beta,\tau_2,\sigma_2}(\kk_2-\xxi)
c_{\beta,\tau_2,\sigma_2}(\kk_2) c_{\alpha,\tau_1,\sigma_1}(\kk_1) \Bigg] \\
 -  \frac{Z_{\text{donor}}e^2}{S}\sum_{\tau,\sigma}\sum_{\kk,\xxi}&\sum_{\alpha=v',c'}\mathcal{V}'(\xxi)c^{\dagger}_{\alpha,\tau,\sigma}(\kk+\xxi)
 c_{\alpha,\tau,\sigma}(\kk) ,
\end{split}
\end{equation}
\begin{equation}\label{eq:Uinter}
\begin{split}
\hat{U}_{\text{inter}} = \frac{e^2}{S}\sum_{\substack{\tau_1,\tau_2\\\sigma_1,\sigma_2}}\sum_{\kk_1,\kk_2,\xxi}\sum_{\alpha=v,c}\sum_{\beta=v',c'}\mathcal{W}(\xxi)&\,c^{\dagger}_{\alpha,\tau_1,\sigma_1}(\kk_1+\xxi)
c^{\dagger}_{\beta,\tau_2,\sigma_2}(\kk_2-\xxi)
 c_{\beta,\tau_2,\sigma_2}(\kk_2) c_{\alpha,\tau_1,\sigma_1}(\kk_1) \\
 -  \frac{Z_{\text{donor}}e^2}{S}\sum_{\tau,\sigma}\sum_{\kk,\xxi}\sum_{\alpha=v,c}\mathcal{W}(\xxi)&\,c^{\dagger}_{\alpha,\tau,\sigma}(\kk+\xxi)
 c_{\alpha,\tau,\sigma}(\kk),
\end{split}
\end{equation}
\end{subequations}
\end{widetext}
where $S$ is the sample area. The donor ion is treated as a dispersionless
scatterer, and is assumed to be present in the MoX${}_2$ (electron)
layer. Henceforth, we assume that a donor yields a single electron to
the TMD and set $Z_{\text{donor}}=1$.

The radiative recombination of electrons and holes is driven
by the light-matter interaction
\begin{equation}\label{eq:lightmatter}
\begin{split}
\hat{H}_{\rm r} =&\frac{e\gamma}{\hbar c}
\sum_{\qq}\sum_{\kk,\tau,\sigma} \sqrt{\frac{4\pi\hbar c }{V q}}
c^{\dagger}_{v,\tau,\sigma}(\kk-\qq_\parallel)c_{c,\tau,\sigma}(\kk)
a_{\tau}^{\dagger}(\qq),
\end{split}
\end{equation}
in the WX${}_2$ layer and an analogous term $\hat{H}_{\rm r}'$ in the
MoX${}_2$ layer.  Here, $\gamma^{(')}$ is given by the in-plane
momentum matrix element between $c^{(')}$ and $v^{(')}$ band states,
evaluated at the $\pm K$ points of the BZ
\cite{kdotp}. $a^{\dagger}_{\tau}(\qq)$ creates a photon of momentum
$\qq$ and in-plane polarization $\tau$, determined by the electron's
valley degree of freedom, where $\tau = +$ ($\tau = -$) represents
right-handed (left-handed) circular polarization. The photon
momentum $\qq = \qq_\parallel + \qq_\perp$ is split into its in-plane
and out-of-plane components, respectively, and $V=SL$, with $L$ the
height of the optical cavity in which the sample is embedded.

Let $| \Psi \rangle$ be an interlayer excitonic eigenstate of the
Hamiltonian $\hat{H}_0+\hat{U}_{\text{intra}}+\hat{U}_{\text{inter}}$
of energy $E_\Psi$. Photon emission through the term $\hat{H}_{\rm r}$
requires the recombining carriers to be in the same TMD layer. This is
allowed by the perturbation $\hat{H}_{\rm t}$, giving the first-order
correction to the wave function,
\begin{equation}\label{eq:first_order}
|\Psi^{(1)}\rangle =
\sum_{n}\frac{\langle n| \hat{H}_{\rm t}|\Psi\rangle}{E_n-E_\Psi}|n\rangle,
\end{equation}
where the sum runs over the eigenstates $|n\rangle$ of
$\hat{H}_{0}+\hat{U}_{\text{intra}}+\hat{U}_{\text{inter}}$, with
energies $E_n$. The resulting rate of radiative recombination is then
given by Fermi's golden rule as
\begin{equation}\label{eq:golden}
\Gamma_{\Psi} = \frac{2\pi}{\hbar} \sum_{f} \left|
\braoketf{f}{\big[\hat{H}_{\rm r}+\hat{H}_{\rm r}' \big]}{\Psi^{(1)}} \right|^2
\delta(E_f-E_\Psi),
\end{equation}
where $\{| f \rangle\}$ is the set of possible final states,
containing one additional photon. As discussed below, the relevant
matrix elements in Eq.\ (\ref{eq:golden}) can be evaluated numerically
in QMC\@.

\section{Recombination of donor-bound interlayer complexes}\label{sec:complexes}

\subsection{Model parameters}

We now discuss the optical emission signatures of the most relevant
donor-bound interlayer excitonic complexes predicted by VMC and DMC
simulations. For concreteness, we will focus on MoSe${}_2$/WSe${}_2$
heterobilayers (X=Se);
parameters relevant to this pair of materials are shown in Table
\ref{tab:mat_parameters}. Furthermore, we assume that the
heterobilayer is encapsulated in bulk hBN, and set the dielectric
constant to $\epsilon=4$.  Our chosen value of $4$ corresponds to the
high-frequency dielectric constant of hBN, which is reasonable as the
exciton binding energy is considerably larger than the highest optical
phonon frequency of hBN\@.  In principle, the anisotropic nature of the
encapsulating hBN supplies an effective dielectric constant
$\bar{\epsilon}=\sqrt{\epsilon_{\parallel}\epsilon_{\perp}}$ and
renormalizes the interlayer distance $d$ by a factor
$\sqrt{\epsilon_{\parallel}/\epsilon_{\perp}}$, where
$\epsilon_{\parallel}$ and $\epsilon_{\perp}$ are the in-plane and
out-of-plane dielectric constants (see Appendix
\ref{sec:multilayer_keldysh}).  However, taking
$\epsilon_{\parallel}(\infty)$ and $\epsilon_{\perp}(\infty)$ from
various sources we find that
$3.1<\sqrt{\epsilon_{\parallel}\epsilon_{\perp}}<4.5$ and
$0.71<\sqrt{\epsilon_{\parallel}/\epsilon_{\perp}}<0.95$
\cite{geick1966normal,plass1997layered,barth1998situ,rumyantsev2001boron}.
This justifies, in part, our use of $\epsilon=4$ and our use of the
unmodified physical layer separation, but as a check of the robustness of our
results, we have also considered a few other dielectric environments for a
restricted set of charge complexes. 

The Hamiltonian of Eq.\ (\ref{eq:H}), without $\hat{H}_{\rm t}$ (i.e., with
charges being fixed in their layers), was solved using DMC for
various numbers of excess electrons and holes, and in the presence of
donor impurities in the MoSe$_2$ layer. Our DMC total energies are
statistically exact: we have not considered any complexes containing
indistinguishable fermions, and therefore the ground-state wave
functions are nodeless, so that no fixed-node error is incurred. The
technical details of our DMC calculations
are given in Appendix \ref{sec:technical}.  Binding energies for free
and impurity-bound excitons and trions, in different dielectric
environments, are reported in Table \ref{table:summ}.
DMC binding energies for a wider range of
  charge-carrier complexes in heterobilayers are reported in Table
  \ref{tab:dmc_dissociations} in Appendix
  \ref{sub:dmc_energies}. A number of donor-bound complexes with up to four
  electrons and two holes are predicted to be stable.  A detailed
  account of the sensitivity of the binding energy of
  D$^0_{c'}$X$_{vc'}$ to our choices of model parameter ($m_{c'}$,
  $m_v$, $r_*$, $r'_*$, $d$, and $\epsilon$) is given in Appendix
  \ref{sec:param_sens}.

\begin{table}[h!]
  \centering
  \caption{Model parameters for MoSe${}_2$ and WSe${}_2$, extracted
    from
    Refs.\ \onlinecite{wang_model,kumar_physicab_2012,berkelbach_prb_2013,kdotp,mostaani_excitonic_prb_2017},
    and the heterobilayer MoSe${}_2$/WSe${}_2$ extracted from
    Refs.\ \onlinecite{band_alignment1,rivera_interlayer,interlayer_distance_hetero}. The
    interlayer gap $\tilde{E}_{\rm g}$ was estimated from the
    luminescence spectrum reported in Ref.\ \onlinecite{rivera_interlayer},
    considering the exciton binding energies of Table
    \ref{table:summ}. From left to right, the single-layer parameters
    are: lattice constant $a$, VB and CB masses $m_v$ and $m_c$,
    screening length $r_*$ in a vacuum environment, and momentum matrix element
    $\gamma$. The heterobilayer parameters are: valence and conduction
    interlayer spacing $\Delta_v$ and $\Delta_c$, interlayer band gap
    $\tilde{E}_{\rm g}$, and interlayer distance
    $d$.
}
  \label{tab:mat_parameters}
  \begin{tabular*}{\columnwidth}{@{\extracolsep{\stretch{1}}}*{7}{lccccc}@{}}
    \hline\hline \\[-1em]
    & $a$ ({\AA}) & $m_v/m_0$ & $m_c/m_0$ & $r_*$ ({\AA}) & $\gamma$ (eV\,{\AA}) \\
    \hline
    MoSe${}_2$           & 3.30      & 0.44        & 0.38  & 39.79   & 2.53  \\
    WSe${}_2$             & 3.29      & 0.34        & 0.29 & 45.11   & 3.17 \\
    \hline\hline
  \end{tabular*} \\[2ex]
  \begin{tabular*}{\columnwidth}{@{\extracolsep{\stretch{1}}}*{7}{lcccc}@{}}
\hline \hline \\[-1em]
    & $\Delta_v$ (eV) & $\Delta_c$ (eV) & $\tilde{E}_{\rm g}$ (eV) & $d$ ({\AA}) \\
    \hline
    MoSe${}_2$/WSe${}_2$& 0.36 & 0.36 & 1.5 &  6.48 \\
    \hline \hline
  \end{tabular*}
\end{table}

The simplest interlayer excitonic complex is a donor-bound
exciton D$^0_{c'}$h$_v$, where D$^0_{c'}$ represents a positive donor ion that has been neutralized by binding an electron from band $c'$, and h$_v$ a hole from band $v$. (When complex
labels appear as subscripts in formulas, we will suppress the $v$ and
$c$ subscripts for clarity.)  DMC simulations predict that this
complex is unbound due to the screening of the interlayer interaction
between holes and the strongly bound neutral donor state D$^0_{c'}$,
whose binding energy is $\mathcal{E}_{\rm D^0}^{\rm b}=-229.03$ meV
(Table \ref{table:summ}). We therefore consider the recombination of a
neutral donor D$^0_{c'}$ with delocalized holes in band $v$.

Adding one more electron we obtain a donor-bound trion. Alternatively,
this complex can be viewed as an interlayer exciton X$_{vc'}$ bound by
a neutral donor D$^0_{c'}$, leading to the notation
D$^0_{c'}$X$_{vc'}$. Remarkably, this larger complex is stable up to $\sim 256$
K\@, with binding
energy ${\cal E}_{\rm D^0X}^{\rm b}\approx 22.52$ meV (Table
\ref{table:summ}) for the most energetically favorable dissociation
channel into a neutral donor D${}^0_{c'}$ and an interlayer exciton
X${}_{vc'}$.

In the following sections we calculate the photoemission rates of
these two complexes using the formalism described in
Sec.\ \ref{sec:model}.

\begin{table}[t!]
  \centering
  \caption{Binding energies ${\cal E}^{\rm b}$ of some charge-carrier
    complexes in a MoSe$_2$ monolayer, a WSe$_2$ monolayer, and a
    MoSe$_2$/WSe$_2$ heterobilayer in different dielectric
    environments including: vacuum on both sides, SiO$_2$ on one side
    and vacuum on the other, bulk hBN on one side and vacuum on the
    other, and bulk hBN on both sides. In the heterobilayer it is
    assumed that the donor ion and electrons occur in the MoSe$_2$
    layer, while the holes are confined in the WSe$_2$ layer.  The
    material parameters are listed in Table
    \ref{tab:mat_parameters}.
    The DMC error bars are everywhere
    smaller than 0.2 meV\@.}
  \label{table:summ}
  \begin{tabular}{lSSSSS}
    \hline\hline

& & \multicolumn{4}{c}{Binding energy (meV)} \\

\raisebox{1.5ex}[0pt]{System} & {\raisebox{1.5ex}[0pt]{$\epsilon$}} &
         {X$_{vc'}$} & {X$^-_{vc'c'}$} & {D$^0_{c'}$} & {D$^0_{c'}$X$_{vc'}$} \\

    \hline

    hBN/MoSe$_2$/hBN  & 4 & 194 & 16.2 & 260 & 21.0 \\

    hBN/WSe$_2$/hBN   & 4 & 160 & 13.6 & 215 & 18.1 \\

\hline

    vac./MoSe$_2$/WSe$_2$/vac.  & 1 & 206 & 6.2 & 540 & 40.3 \\

    SiO$_2$/MoSe$_2$/WSe$_2$/vac. & 2.45  & 123 & 5.1 & 329 & 30.1 \\

    hBN/MoSe$_2$/WSe$_2$/vac. & 2.5 & 121 & 5.2 & 324 & 29.9 \\

   hBN/MoSe$_2$/WSe$_2$/hBN & 4 & 84.2 & 4.1 & 229 & 22.5 \\

    \hline\hline
  \end{tabular}
\end{table}

\subsection{D$^0_{c'}$h$_{v}$: Neutral donor and free hole}\label{sec:D0h}

The initial state for the recombination process of a neutral donor and
a free hole is given in second quantization by
\begin{equation}\label{eq:D0}
|{\rm D}^0; \kk_{\rm h} \rangle =
\frac{1}{\sqrt{S}}\sum_{\kk}\tilde{\chi}_{\kk}\,c^{\dagger}_{c',\tau',\sigma'}(\kk)c_{v,\tau,\sigma}(\kk_{\rm
  h})|\Omega\rangle,
\end{equation}
where $\tilde{\chi}_{\kk}=\int \chi(\rr) e^{-i\kk\cdot \rr} \, d^2r$
is the Fourier transform of the donor-atom wave function centered at
the donor site. Relative to the neutral vacuum, the state's energy can
be written as $E_{\rm D^0}(\kk_{\rm h}) = E_{c'}(0)-E_{v}(\kk_{\rm h})
- {\cal E}_{\rm D^0}^{\rm b}$, with ${\cal E}_{\rm D^0}^{\rm b}$ the
binding energy.

In the close-alignment limit and in the absence of
intervalley scattering, the complex described by Eq.\ (\ref{eq:D0})
can decay through radiative recombination only if
$\tau'=\tau$. Furthermore, spin-valley locking \cite{wang_model} and
the known band ordering of MoSe${}_2$ and WSe${}_2$ monolayers
\cite{kdotp} further require that $\sigma = \sigma'$. Considering
single-photon final states of the form
$\ket{f}=a^{\dagger}_{\tau}(\qq)\ket{\Omega}$, with polarization
determined by the valley quantum number, and assuming a small twist
angle $\theta \approx 0^\circ$, Eqs.\ (\ref{eq:first_order}) and
(\ref{eq:golden}) give the radiative decay rate
\begin{equation}\label{eq:d0h_lesser}
\begin{split}
\Gamma_{\rm D^0h}^{<}= &\, \frac{4 \tilde{E}_{\rm g}\abs{F(\rr_0)}^2}{\hbar}\frac{e^2}{\hbar c}\left[
\frac{t_{vv}\gamma'}{\hbar c\Delta_v}-\frac{t_{cc}\gamma}{\hbar c(\Delta_c+{\cal E}^{\rm b}_{\rm D^0})}
\right]^2
\\
&{} \times \left|
\int d^2r\, e^{i\Delta \KK\cdot \rr }\chi(\rr)\right|^2n_{\rm h},
\end{split}
\end{equation}
where $n_{\rm h}$ is the hole density of the sample, and the stacking-dependent function $F(\rr_0)=1+\exp{-i\GG_1\cdot\rr_0}+\exp{-i\GG_2\cdot\rr_0}$ is introduced by the momentum-conservation rule in Eq.\ \eqref{eq:hopping} (see also Fig.\ \ref{fig:moire} in Appendix \ref{sec:amp}). To evaluate
Eq.\ (\ref{eq:d0h_lesser}), we obtain the wave function $\chi(\rr)$ of
the donor-bound electron by solving the two-body problem with a
finite-elements method, as detailed in Appendix \ref{sec:finite}. Note
that a finite amplitude for radiative recombination depends critically
on the electron-hole asymmetry, and on having different tunneling
strengths between the CBs and the VBs of the two layers, owing to the
symmetry properties of the band states.

\subsection{D$^0_{c'}$X$_{vc'}$: Donor-bound interlayer trion}
\label{sec:d0ctvct}

As discussed above, a donor-bound trion D$^0_{c'}$X$_{vc'}$ can be
viewed as an interlayer exciton bound to a neutral donor
ion. Defining the interlayer exciton X$_{vc'}$ and D$_{c'}^0$
energies as $E_{\rm X} = E_{c'}(0)-E_{v}(0) - {\cal E}_{\rm X}^{\rm b}$ and $E_{\rm D^0}
= E_{c'}(0) - {\cal E}_{\rm D^0}^{\rm b}$, respectively, the energy of a
D$_{c'}^0$X$_{vc'}$ complex can be expressed as $E_{\rm D^0X} = E_{\rm D^0} +
E_{\rm X} - {\cal E}_{\rm D^0X}^{\rm b}$, where ${\cal E}_{\rm D^0X}^{\rm b}$ is the binding
energy defined with respect to the most favorable dissociation channel into D${}_{c'}^{0}+$X${}_{vc'}$. The corresponding eigenstate is given by
\begin{equation}\label{eq:D0X}
\begin{split}
|{\rm D^0X} \rangle = &\,
\frac{1}{S^{3/2}}\sum_{\kk_{\rm h},\kk_1,\kk_2}\tilde{\Phi}_{\kk_{\rm h},\kk_1,\kk_2}
\\
&\times
c^{\dagger}_{c',\tau',\sigma}(\kk_1)
c^{\dagger}_{c',-\tau',-\sigma'}(\kk_2)
c_{v,\tau,\sigma}(\kk_{\rm h})|\Omega\rangle,
\end{split}
\end{equation}
with its two electrons belonging to opposite valleys, thus minimizing
their mutual repulsion [see Eqs.\ (\ref{eq:exact_int_11}) and
  (\ref{eq:exact_int_22})]. In this case, we consider decay into
states of the form
$\ket{f}=a_{\tau}^\dagger(\qq)|\mathrm{D}^0\rangle$, which are
energetically favorable given the large binding energies of
$\mathrm{D}_{c'}^0$ bound states. The corresponding radiative rate for
close interlayer alignment is given by
\begin{equation}
\begin{split}
&\Gamma_{\rm D^0X}^{<}\approx \frac{4 \tilde{E}_{\rm
      g}}{\hbar}\frac{e^2}{\hbar c}\abs{F(\rr_0)}^2\\
      &\times\left| \int d^2r \int
  d^2r'\,e^{i\Delta \KK\cdot \rr}
  \chi^*(\rr')\Phi(\rr,\rr,\rr')\right|^2 \\ &\times\left[
    \frac{t_{vv}\gamma'}{\hbar c(\Delta_v+{\cal E}^{\rm b}_{\rm
        D^0X}+{\cal E}^{\rm b}_{\rm X})}-\frac{t_{cc}\gamma}{\hbar
      c(\Delta_c+{\cal E}^{\rm b}_{\rm D^0X}+{\cal E}^{\rm b}_{\rm
        X})} \right]^2.
\end{split}
\label{eq:gd0xsmall}
\end{equation}

The donor atom in the final state can be in its ground state, or in
any excited state allowed by angular momentum conservation. This
constitutes a series of radiative subchannels, and in principle
results in a series of lines with energies determined by the donor
atom spectrum. The main subchannel, corresponding to the ground state
$\chi_{\rm 1s}(\rr)$, produces the main emission line at
$E_{*}=\tilde{E}_g-(\mathcal{E}_{\mathrm{D^0X}}^{\mathrm{b}}+\mathcal{E}_{\mathrm{X}}^{\mathrm{b}})$. The
first radially symmetric excited state, $\chi_{\rm 2s}(\rr)$, will
produce an additional line $\sim 167$ meV
above the main line. The overlap integrals
between the ground-state donor-bound trion and the 1$s$ and 2$s$ neutral
donor states were evaluated using VMC, and the latter was found to be
two orders of magnitude smaller.
We conclude that excited states can be neglected, and
henceforth only the $1s$ subchannel will be considered. In the case of $\Delta K = 0$, the integral in Eq.\ (\ref{eq:gd0xsmall}) is given by
$\left|
\int d^2r \int d^2r'\,\chi^*(\rr')\Phi(\rr,\rr,\rr')\right|^2=1.47$ (see Appendix \ref{sub:overlaps} for details).

To summarize Sec.\ \ref{sec:complexes}, Fig.\ \ref{fig:figure1}(b)
shows the radiative rates of D$^0_{c'}$h$_{v}$ and D$^0_{c'}$X$_{vc'}$ in an
hBN/MoSe$_2$/WSe$_2$/hBN heterostructure, for small twist angles and using the maximum value of $\abs{F(0)}^2=9$. Alternatively, we may average this function within the unit cell, leading to $\langle\abs{F(\rr_0)}^2 \rangle\approx 3$. The
large-angle asymptotic behavior of the radiative rate shown in
Fig.\ \ref{fig:figure1}(b) is discussed next.

\section{Asymptotic behavior for large interlayer twist angles}\label{sec:asymptotics}
\begin{figure}[t!]
  \centering
  \includegraphics[width=0.40\textwidth]{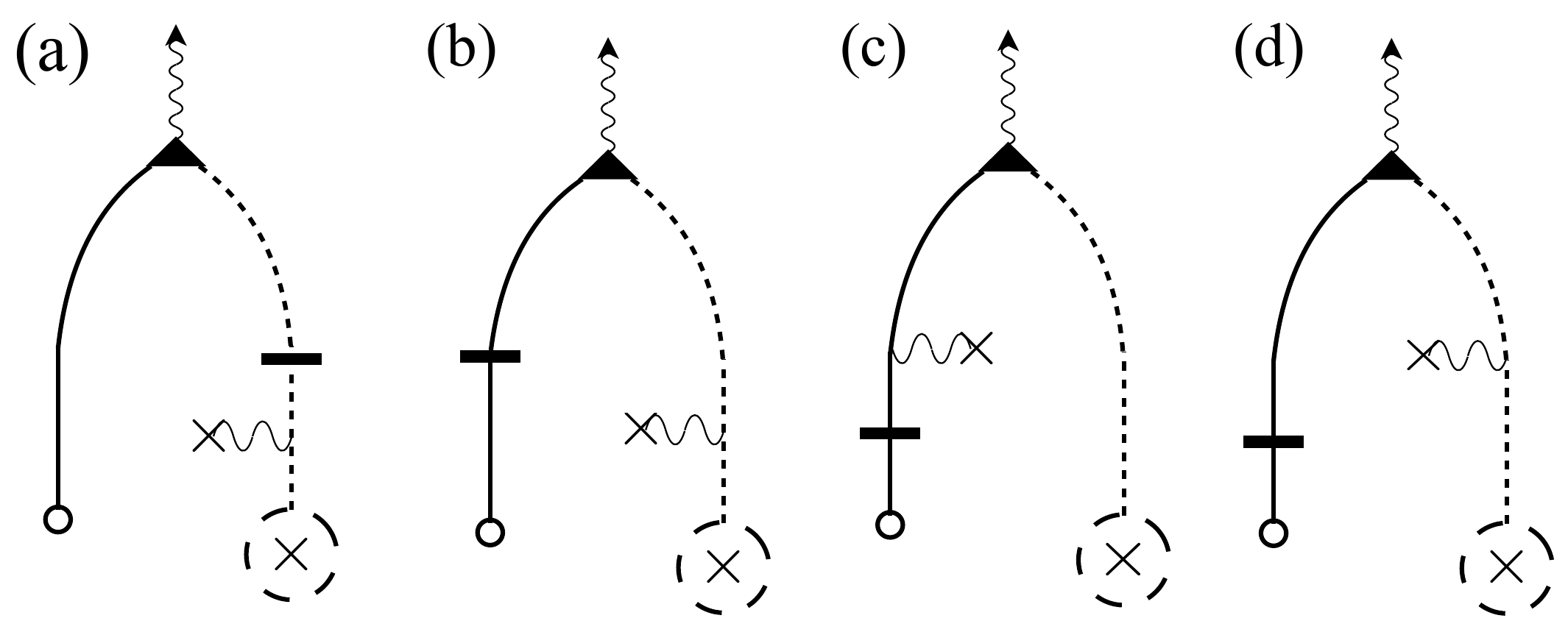}
  \caption{Diagrams for the radiative recombination of neutral donors
    D$^0_{c'}$ with free holes h${}_v$. The solid (dashed) line represents a free
    hole (electron); the donor impurity center is represented by a
    ``$\times$'' symbol, and the D$^0_{c'}$ state by ``$\times$'' in a
    dashed circle. Horizontal lines correspond to interlayer
    tunneling, wavy lines to Coulomb scattering, and the triangular
    vertex represents radiative recombination.}
  \label{fig:d0h_diagrams}
\end{figure}

To estimate the quenching of radiative decay as the misalignment angle grows, we evaluate the asymptotic behavior of the radiative rate for large valley mismatch
$|\Delta\KK| \gtrsim 1/a_0^*$ from a
perturbative treatment of the short-range interactions
(\ref{eq:short_range_intra}). In this regime, the rate of radiative decay of intralayer complexes is determined by the tail of the momentum-space wavefunction extending toward the opposite layer valley, and which is governed by the large-momentum portion of the interaction term \eqref{eq:Uintra}. Thus, we formally split $\hat{U}_{\rm intra}=\hat{U}_{\rm intra}^<+\hat{U}_{\rm intra}^>$ and $\hat{U}_{\rm inter}=\hat{U}_{\rm inter}^<+\hat{U}_{\rm inter}^>$, where ``large'' ($>$) momentum corresponds to wave vectors $\gtrsim 1/a_0^*$. Let $| \Psi_0 \rangle$ be an
excitonic state of energy $E_\Psi^0$, of the Hamiltonian
\begin{equation}\label{eq:HLR}
  \hat{H}_{\text{LR}} = \hat{H}_0 + \hat{U}_{\text{intra}}^{<} + \hat{U}_{\text{inter}}^{<},
\end{equation}
containing the long-range approximation to the
carrier-carrier and donor-carrier interaction. The interactions
$\hat{U}_{\text{intra}}^{<}$ and $\hat{U}_{\text{inter}}^{<}$ are
given by the expressions (\ref{eq:Uintra}) and (\ref{eq:Uinter}),
respectively, with the substitutions $\mathcal{V}^{(')}(\xxi)
\longrightarrow \mathcal{V}^{(')}_{<}(\xxi)$ and $\mathcal{W}(\xxi)
\longrightarrow \mathcal{W}_{<}(\xxi)$ [see
  Eqs.\ (\ref{eq:keldysh_form_nn}) and
  (\ref{eq:keldysh_form_12})]. The state $| \Psi_0\rangle$ is
perturbed by the interlayer tunneling term $\hat{H}_{\rm t}$, as well
as the short-range interaction $\hat{U}_{\text{intra}}^{>}$, obtained
by substituting $\mathcal{V}^{(')}(\xxi) \longrightarrow
\mathcal{V}^{(')}_{>}(\xxi)$ in Eq.\ (\ref{eq:Uintra}) [see
  Eq.\ (\ref{eq:short_range_intra})]. As shown in Eq.\ \eqref{eq:short_range_intra}, these terms are inversely proportional to the square of a large wave number, and thus may be treated perturbatively. Furthermore, the
short-range interlayer term is exponentially suppressed, and can be
ignored altogether. As a consequence, short-range impurity scattering
can take place exclusively in the electron layer, where the impurity centers are located (see diagrams of
Fig.\ \ref{fig:d0h_diagrams}).

\begin{figure}[]
  \centering
  \includegraphics[width=0.45\textwidth]{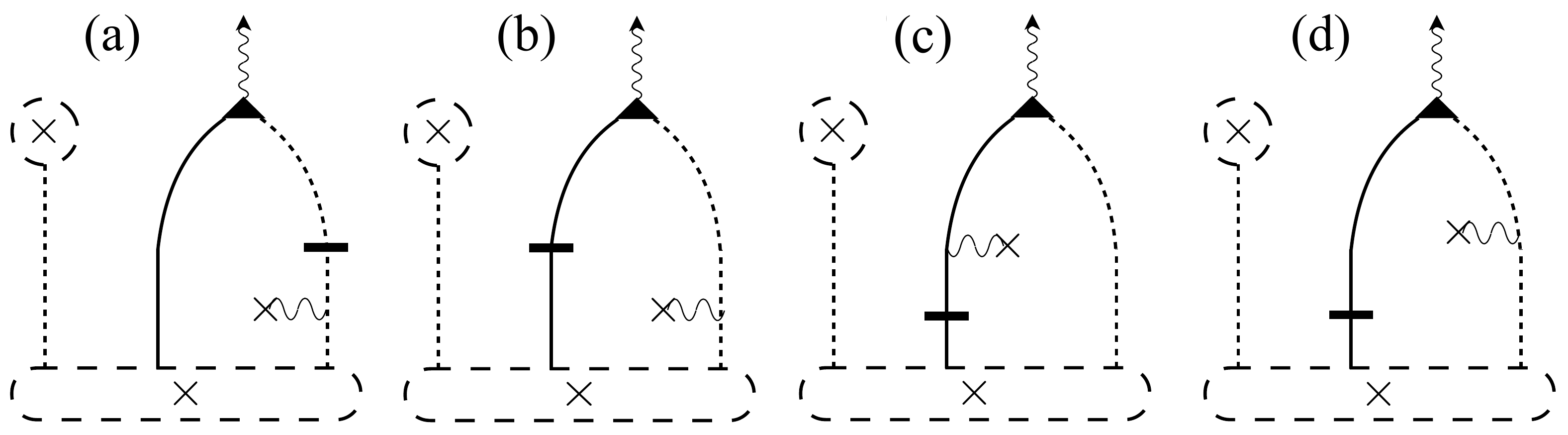}
  \caption{Diagrams for the first radiative recombination channel of
    the D$^0_{c'}$X$_{vc'}$ complex. The bound hole recombines with the
    electron from the nearest valley in the opposite layer, assisted by short-range Coulomb interactions with the
    donor impurity. The remaining electron stays bound to the impurity
    center, forming a neutral donor atom.}
  \label{fig:d0X_diagrams1}
\end{figure}

In second-order perturbation theory, the correction to the wave function relevant for
photon emission is given by
\begin{equation}\label{eq:second_order}
|\Psi_0^{(2)}\rangle = \sum_{m,n}\frac{\langle n| [\hat{H}_\mathrm{t} +
    \hat{U}_{\text{intra}}^{>} ]|m\rangle\langle m| [\hat{H}_\mathrm{t} +
    \hat{U}_{\text{intra}}^{>}
  ]|\Psi_0\rangle}{(E_m^0-E_\Psi^0)(E_n^0-E_\Psi^0)}|n\rangle,
\end{equation}
where the sums run over the eigenstates $|n\rangle$ of
$\hat{H}_{\text{LR}}$, with energies $E_n^0$. Introducing the
light-matter interaction [Eq.\ (\ref{eq:lightmatter})], we focus on the
diagrams of Fig.\ \ref{fig:d0h_diagrams} for the D${}_{c'}^0$h$_{v}$
complex, and those of Fig.\ \ref{fig:d0X_diagrams1} for
D${}_{c'}^0$X${}_{vc'}$.

In general, all diagrams must be considered when evaluating the
radiative decay rate. For simplicity, however, we assume that the CB
and VB spacings remain the largest scales in the problem, such that
$\tfrac{\hbar^2\Delta K^2}{2m_{\alpha}} \ll \Delta_c,\,\Delta_v$. In
this approximation, two out of the four diagrams for
$\mathrm{D}_{c'}^0\mathrm{h}_{v}$ radiative decay cancel out
approximately, leaving only the contributions from the diagrams of
Figs.\ \ref{fig:d0h_diagrams}(a) and \ref{fig:d0h_diagrams}(b) (see Appendix
\ref{sec:amp}). The resulting radiative decay rate for
D$^0_{c'}$h$_{v}$ in the
large twist angle ($>$) limit is
\begin{equation}
\begin{split}
\Gamma^>_{\rm D^0h}\approx&\,
\frac{64 \pi^2 e^4 \tilde{E}_{\rm g}}{\hbar \epsilon^2 r_{*}'{}^2\Delta K^4}
\frac{e^2}{\hbar c}
\left[ \frac{m_{c'}}{\hbar^2\Delta K^2}\right]^2
\left[\frac{t_{cc}\gamma}{\hbar c \Delta_c}-\frac{t_{vv}\gamma'}{\hbar c \Delta_v}\right]^2
\\
&\times
|\chi_0(0)|^2\abs{F(\rr_0)}^2n_{\rm h},
\end{split}
\label{eq:d0h}
\end{equation}
where the emitted photon energy is given by $E_* = \tilde{E}_\mathrm{g}-{\cal
E}^{\rm b}_{\rm D^0}$.
Finally, $\chi_0(\rr)$ is the $\mathrm{D}_{c'}^0$ wave
function obtained from the Keldysh approximation Hamiltonian
$\hat{H}_{\text{LR}}$, not to be confused with the full bilayer interaction bound state $\chi(\rr)$. As before, we evaluate the wave function using the
finite-element method, and obtain
$|\chi_0(0)|^2=2.678\times 10^{-3}${\AA}$^{-2}$ (Appendix \ref{sec:finite}). 
We point out that evaluating the wave function $\chi_0(\rr)$ with the Keldysh potential ignores the formal wave vector cutoff that defines Eq.\ \eqref{eq:HLR}. That is, this solution considers short range interactions within the Keldysh approximation, which, as discussed in Sec.\ \ref{sec:model_bilayer_int}, overestimate the screening length. Nonetheless, this approximation mainly affects the fast oscillating (large momentum) part of the wave function, whereas Eq.\ \eqref{eq:d0h}, and Eqs.\ \eqref{eq:d0x1} and \eqref{eq:d0x2} below, only depend on the smooth, small momentum part. The error incurred by this approximation is proportional to the perturbation squared, and thus beyond our first-order approximation.

With the perturbation $\hat{U}_{\text{intra}}^{>}$, there are two
possible channels for radiative recombination of the
$\mathrm{D}_{c'}^0\mathrm{X}_{vc'}$ complex, resulting in different
final states, and thus two separate lines in the PL spectrum. The
first process involves one of the electrons and the hole scattering
from the donor impurity and subsequently recombining, emitting a
photon and leaving behind a neutral donor as the final state. This is
analogous to the decay process considered in Sec.\ \ref{sec:d0ctvct},
and the corresponding diagrams are shown in
Fig.\ \ref{fig:d0X_diagrams1}. Similarly to the D$_{c'}^0$h$_v$ complex
case, the leading approximation to the amplitude is the sum of two
diagrams, giving a radiative rate
\begin{equation}
\begin{split}
\Gamma^>_{\rm D^0X} \approx &\, \frac{64\pi^2 e^4 \tilde{E}_\mathrm{g}}{\hbar \epsilon^2r_{*}'{}^2 \Delta K^4}\frac{e^2}{\hbar c} \left[ \frac{m_{c'}}{\hbar^2\Delta K^2}\right]^2
\left[\frac{t_{cc}\gamma}{\hbar c \Delta_c}-\frac{t_{vv}\gamma'}{\hbar c \Delta_v}\right]^2
\\
&\times\left|
\int d^2 r\, \chi_0^* (\rr)\Phi_0(0,0,\rr)
\right|^2\abs{F(\rr_0)}^2,
\end{split}
\label{eq:d0x1}
\end{equation}
where the emitted photon energy is given by $E_* =
\tilde{E}_\mathrm{g}-({\cal E}^{\rm b}_{\rm D^0X}+{\cal E}^{\rm b}_{\rm X})$, and $\Phi_0(\rr_{\rm h},\rr_{\rm e},\rr_{\rm e'})$ is the D${}_{c'}^0$X${}_{vc'}$ wave function in the Keldysh approximation.

\begin{figure}[t!]
  \centering
  \includegraphics[width=0.95\columnwidth]{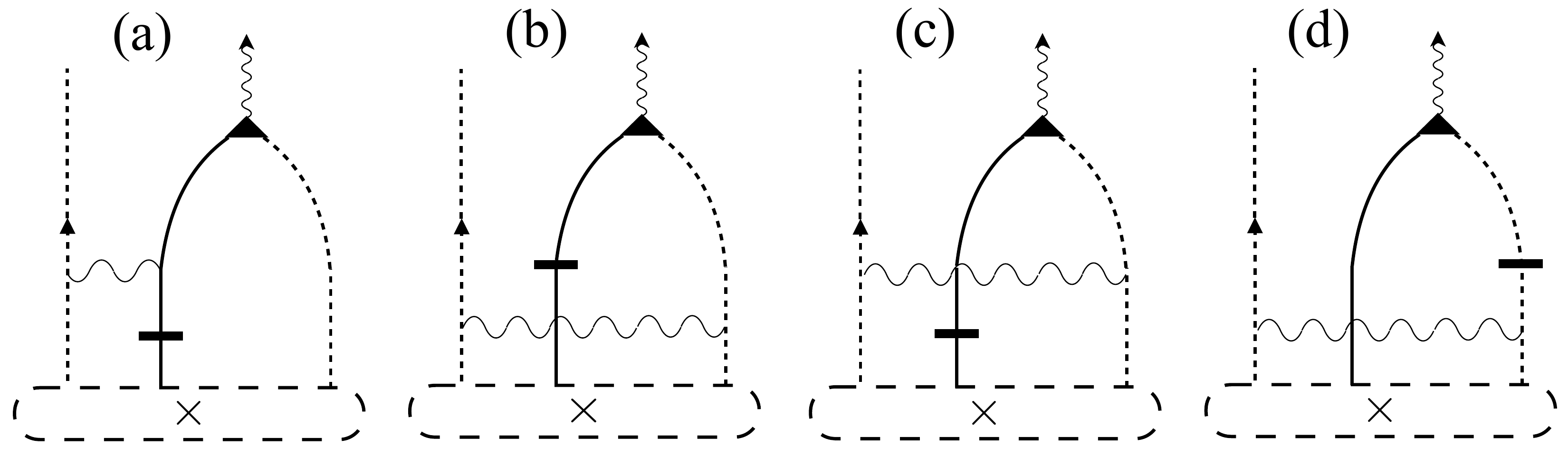}
  \caption{Diagrams for the second radiative recombination channel of
    the D$^0_{c'}$X$_{vc'}$ complex. The bound hole recombines with the
    electron from the nearest valley in the opposite layer, assisted by short-range Coulomb interactions with the
    second electron, at the far valley. The latter recoils and unbinds from the donor
    impurity.}
  \label{fig:d0X_diagrams2}
\end{figure}

A second radiative decay process is possible, where the recombining
electron and hole scatter with the second electron, at the far
valley. The latter electron recoils and is unbound from the impurity,
taking some amount of kinetic energy and producing a shift in the
emission line. The corresponding diagrams are shown in
Fig.\ \ref{fig:d0X_diagrams2}, and give a recombination rate

\begin{equation}\label{eq:d0x2}
\begin{split}
\Gamma^{'>}_{\rm D^0X} = &\, \frac{48\pi^2 e^4 \tilde{E}_\mathrm{g}}{\hbar \epsilon^2 r_{*}'{}^2\Delta K^4}\frac{e^2}{\hbar c}
\left[\frac{m_{c'}}{\hbar^2\Delta K^2} \right]^2\left[
\frac{t_{cc}\gamma}{\hbar c\Delta_c}-\frac{t_{vv}\gamma'}{\hbar c \Delta_v}\right]^2
\\
&\times\int d^2r \,\left|\Phi_0(\rr,\rr,\rr) \right|^2.
\end{split}
\end{equation}
The photon energy in this case is given by $E_* = \tilde{E}_{\rm g} -
{\cal E}^{\rm b}_{\rm D^0}-{\cal E}^{\rm b}_{\rm D^0X}-{\cal E}^{\rm
  b}_{\rm X}-\frac{\hbar^2\Delta K^2}{2m_{c'}}$, and the corresponding
line in the PL spectrum is red shifted with respect to that of the
first channel by $\sim 100$ meV\@. Notice the absence of the interference
term $\abs{F(\rr_0)}^2$. For this decay channel, the three
tunneling processes encoded in Eq.\ \eqref{eq:hopping} result in different momenta
for the recoiling electron, and consequently in three distinguishable final states that cannot interfere.

The overlap integrals between the initial- and final-state
wave functions given in Eqs.\ (\ref{eq:d0x1}) and (\ref{eq:d0x2}) were
evaluated in VMC for the Hamiltonian $\hat{H}_{\text{LR}}$. We obtain
$\left|\int d^2 r\, \chi_0^* (\rr)\Phi_0(0,0,\rr)\right|^2=6.94\times
10^{-7}$ {\AA}$^{-4}$, and $\int
d^2r\,\left|\Phi_0(\rr,\rr,\rr)\right|^2=3.22\times 10^{-7}$ $\Ams^{-4}$,
respectively (see Appendix \ref{sub:overlaps}).

Equations (\ref{eq:d0h}), (\ref{eq:d0x1}), and (\ref{eq:d0x2}) show
that the radiative channels considered for the two complexes decay
with the interlayer twist angle as $\theta^{-8}$, in the limit $\Delta
K \gg 1/r_*,1/r_*'$. This is shown in Fig.\ \ref{fig:figure1}(b) for
angles larger than $6^\circ$. Our analysis indicates that, even in the
case of localized impurity-bound states, the observation of
photoluminescence from interlayer excitonic complexes in TMD bilayers
requires near perfect alignment between the two layers.

\section{Phonon-assisted recombination}\label{sec:e_ph_effects}

Electron-phonon (e-ph) interactions introduce yet another channel for
radiative recombination.  Similarly to the electron recoil process
discussed above, when phonons are emitted during the recombination of
a given complex, they absorb part of the energy and produce a
red shifted replica in the PL spectrum. The following analysis is
carried out in terms of the VMC wave
functions $| \Psi \rangle$ discussed in Sec.\ \ref{sec:complexes},
evaluated with the exact bilayer interactions
$\mathcal{V}^{(')}(\xxi)$ and $\mathcal{W}(\xxi)$.

The e-ph interaction Hamiltonian is given by
\begin{equation}\label{eq:e-p}
\begin{split}
\hat{H}_{\text{e-ph}}=\sum_{\alpha=v,c}\sum_{\tau,\sigma}\sum_{{ \kk},{ \qq},\nu}& \frac{g_{\nu,\alpha}(\qq)}{\sqrt{S}}(b^{\dagger}_{h,\nu,-\qq}+b_{h,\nu,\qq})\\
&\times c^{\dagger}_{\alpha,\tau,\sigma}(\kk+\qq)c_{\alpha,\tau,\sigma}(\kk)\\
+\sum_{\alpha=v',c'}\sum_{\tau,\sigma}\sum_{{ \kk},{ \qq},\nu}& \frac{g_{\nu,\alpha}(\qq)}{\sqrt{S}}(b^{\dagger}_{e,\nu,-\qq}+b_{e,\nu,\qq})\\
&\times c^{\dagger}_{\alpha,\tau,\sigma}(\kk+\qq)c_{\alpha,\tau,\sigma}(\kk),
\end{split}
\end{equation}
where $b^{\dagger}_{\Lambda,\nu,{ \qq}}$ ($b_{\Lambda,\nu,{ \qq}}$) is
the creation (annihilation) operator for a phonon of momentum $\qq$
and mode $\nu$ in the electron ($\Lambda={}$e) or hole ($\Lambda={}$h)
layer, which couples to an electron in band $\alpha=c',v',c,v$ with
strength $g_{\nu,\alpha}(\qq)$.

We consider the longitudinal optical ($\nu=\text{LO}$), homopolar
($\nu=\text{HP}$), and longitudinal acoustic ($\nu=\text{LA}$) phonon
modes allowed by the lattice symmetry.
The e-ph couplings are given by
\begin{equation}
\begin{split}
&g_{\text{LO},\alpha}(\qq) = \frac{1}{A}\sqrt{\frac{\hbar}{2 \rho (M_{\rm r}/M) \omega_{\text{LO}}}}\frac{2\pi Z_\alpha e^2}{1+qr_{*}},
\\
&g_{\text{HP},\alpha}(\qq) = \sqrt{\frac{\hbar}{2 \rho \omega_{\text{HP}}}}D_{\alpha},
\\
&g_{\text{LA},\alpha}(\qq) = \sqrt{\frac{\hbar}{2 \rho \omega_{\text{LA}}}} \Xi_{\alpha}\, q, 
\\
\end{split}
\end{equation}
where $\rho$ is the mass density, $M_{\rm r}$ is the
metal-and-two-chalcogen system reduced mass, $M$ is the total mass of
the unit cell, and $A$ is the unit-cell area of the corresponding TMD
layer. $\omega_{\nu}$ is the phonon frequency, which we approximate as
a constant for the optical modes, and as
$\omega_{\text{LA}}=c_{\text{LA}}\,q$ for the LA mode, with
$c_{\text{LA}}$ being the sound velocity.  $Z$ is the Born effective
charge, $r_{*}$ is the screening length, and $D_\alpha$ and
$\Xi_\alpha$ are the deformation potentials of the optical and
acoustic modes, respectively.
The various parameters are taken from
Refs.\ [\onlinecite{danovich_phonons,chinese_phonons,chinese_phonons2,danish_phonons}],
and summarized in Table \ref{tab:phonon_params}. We focus on the
low-temperature limit, where phonon occupation is low and phonon absorption
can be neglected.

\begin{table*}
  \caption{Electron-phonon coupling parameters for LO, HP, and LA
    phonon modes. $\omega_{\rm LO}$ and $\omega_{\rm HP}$ are the LO-
    and HP-mode frequencies, $c_{\rm LA}$ is the speed of sound for
    the LA mode, $\rho$ is the mass density, $D_\alpha$ and
    $\Xi_\alpha$ are the deformation potentials of the optical and
    acoustic modes, respectively, $M_{\rm r}/M$ is the ratio of
    the metal-and-two-chalcogen system reduced mass to the total mass
    of the unit cell, and $Z$ is the Born effective charge.\label{tab:phonon_params}}
\begin{tabular}{lcccccccccS}
  \hline \hline
    & $\hbar\omega_{\text{LO}}$ (meV) & $\hbar\omega_{\text{HP}}$ (meV) &  $c_{\text{LA}}$ (cm/s) & $\rho$ (g/cm$^2$) & $D_c$ (eV/{\AA}) & $D_v$ (eV/{\AA})
    & $\Xi_c$ (eV) & $\Xi_v$ (eV) &  $M_{\rm r}/M$    & $Z$  \\
\hline
  MoSe${}_2$           & 37       &    30         & $4.8\times 10^5$   & $4.5\times 10^{-7}$ & 5.2   & 4.9 & 3.4  & 2.8 & 0.235 & 1.8  \\
  WSe${}_2$          & 31 & 31  & $4.4\times 10^5$     &  $6.1\times 10^{-7}$ &  2.3 & 3.1   &    3.2       & 2.1 & 0.249 & 1.08  \\ \hline \hline
\end{tabular}
\end{table*}

\begin{figure}[t!]
  \centering
  \includegraphics[width=0.37\textwidth]{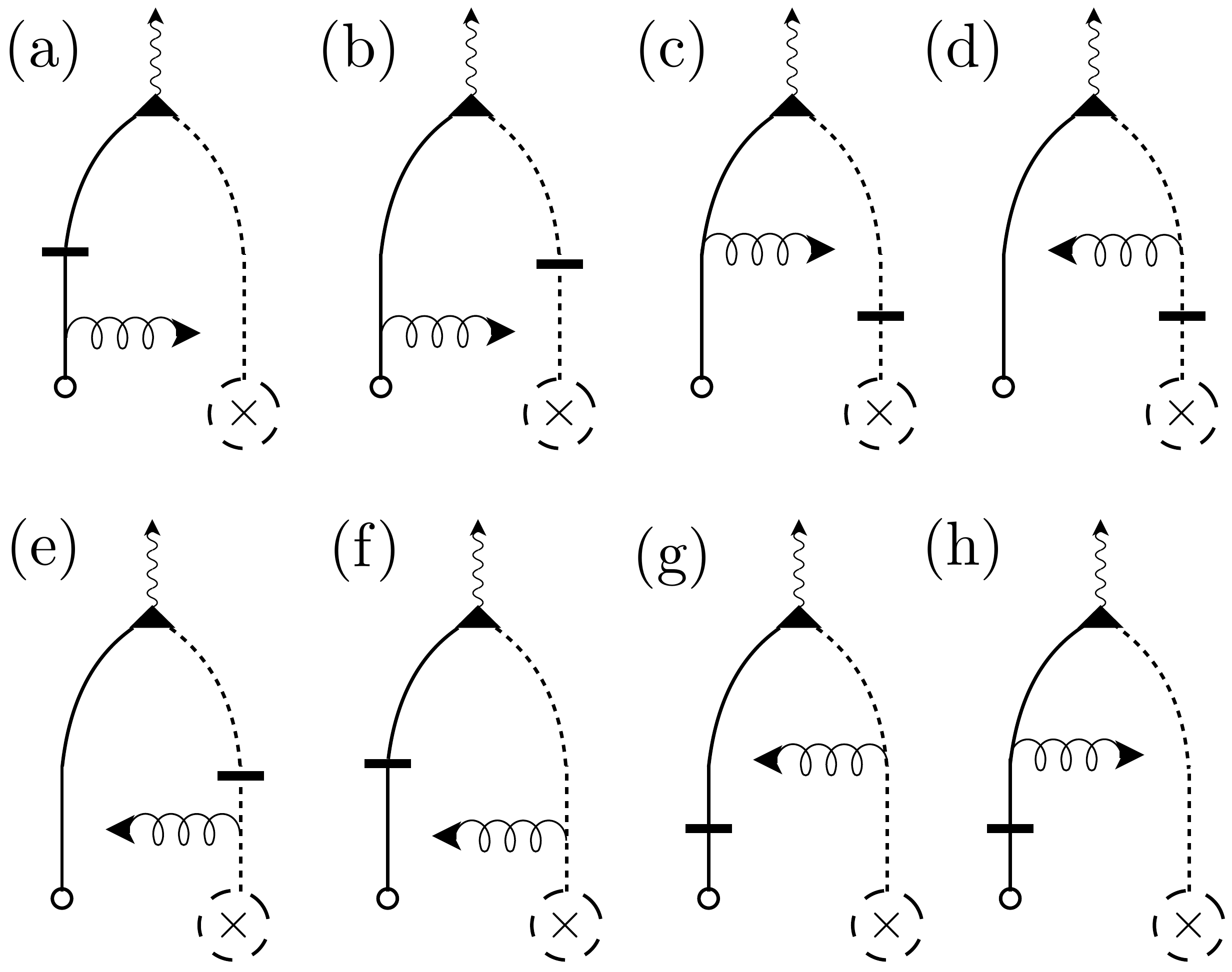}
  \caption{Diagrams for the radiative recombination of the D$^0_{c'}$h$_{v}$
    complex with phonon scattering. The top four diagrams correspond
    to phonon emission in the WSe$_2$ layer and the bottom four
    diagrams correspond to phonon emission in the MoSe$_2$
    layer. }
  \label{fig:d0h_diagrams_phonons}
\end{figure}

\begin{figure}[t!]
  \centering
  \includegraphics[width=0.47\textwidth]{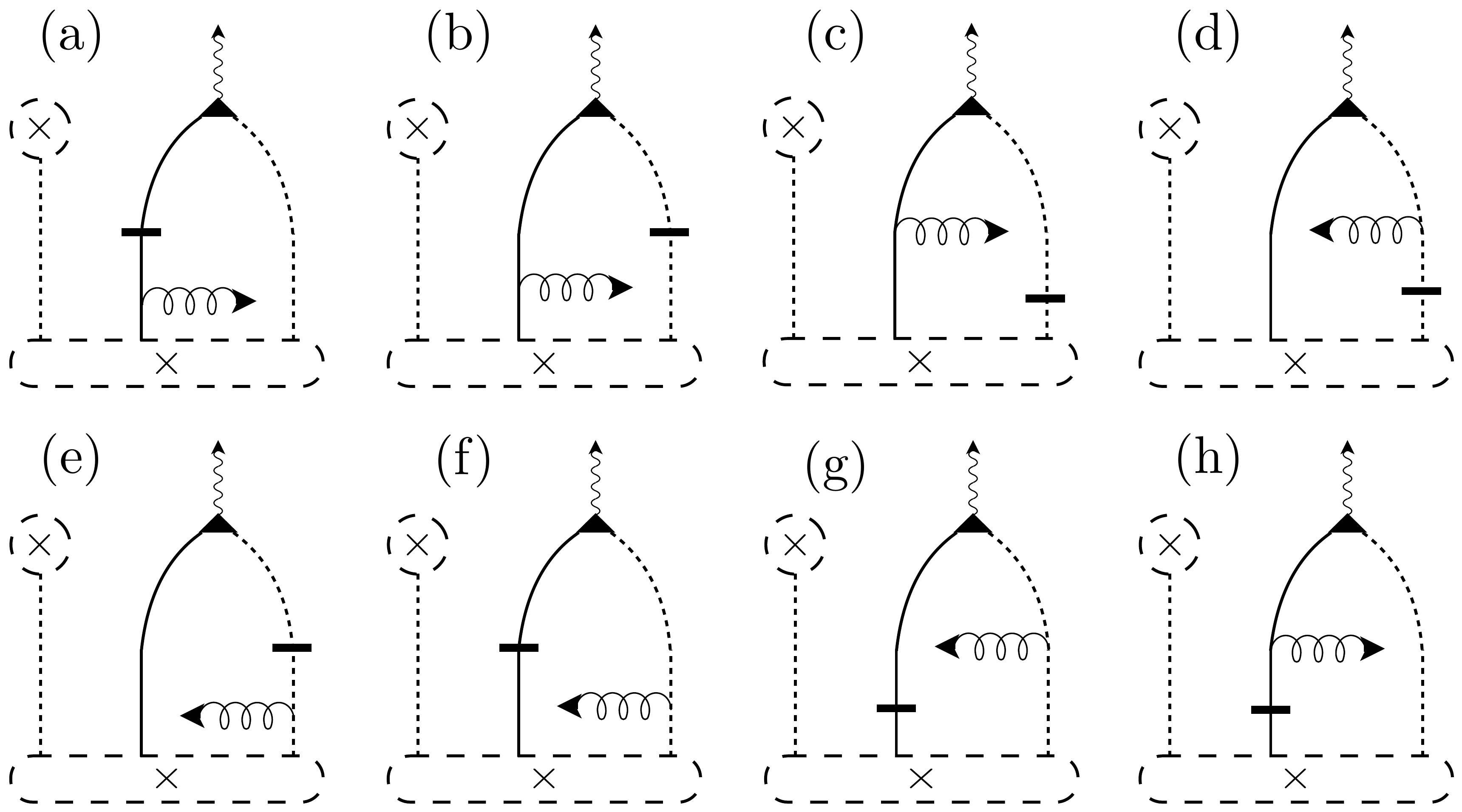}
  \caption{Diagrams for the radiative recombination of the D$^0_{c'}$X$_{vc'}$
    complex with phonon scattering and D$^0_{c'}$ in the final state. The
    top four diagrams correspond to phonon emission in the WSe$_2$
    layer and the bottom four diagrams correspond to phonon emission
    in the MoSe$_2$ layer.}
  \label{fig:d0X_diagrams_phonons}
\end{figure}

Perturbative corrections to the interlayer excitonic state $|\Psi
\rangle$ by the interlayer hopping and e-ph interactions
are given by
\begin{equation}\label{eq:second_order_ph}
|\Psi^{(2)}\rangle =
\sum_{m,n}\frac{\langle n| [\hat{H}_{\rm t} + \hat{H}_{\text{e-ph}} ]|m\rangle\langle m| [\hat{H}_{\rm t} + \hat{H}_{\text{e-ph}} ]|\Psi\rangle}{(E_m-E_\Psi)(E_n-E_\Psi)}|n\rangle.
\end{equation}
The relevant diagrams for radiative recombination with phonon emission are
shown in Figs.\ \ref{fig:d0h_diagrams_phonons} and
\ref{fig:d0X_diagrams_phonons} for D$^0_{c'}$h$_v$ and
D$^0_{c'}$X$_{vc'}$, respectively.  In both figures, panels (a)--(d)
correspond to single-phonon emission in the hole layer (WSe$_2$),
whereas panels (e)--(h) correspond to single-phonon emission in the
electron layer (MoSe$_2$). Although, in principle, the two sets of
diagrams give separate lines at energies determined by the phonon
energy in each layer, the parameters reported in Table
\ref{tab:phonon_params} show that these lines are within only a few
meV of each other. For simplicity, we assume that the two layers have
the same optical-phonon energies and the same acoustic-phonon sound velocities,
producing a single line in the PL spectrum. The resulting radiative
rates are given in the limit of large twist angle $(>)$ by (Appendix
\ref{app:phonon_assisted})
\begin{subequations}
\begin{equation}\label{eq:G_d0h_phonon_gtr}
\begin{split}
&\Gamma^{>,\nu}_{\rm D^0h} \approx \frac{48\tilde{E}_{\rm g}}{\hbar}\frac{e^2}{\hbar c}\left[\frac{\gamma't_{vv}}{\hbar c \Delta_v}-\frac{\gamma t_{cc}}{\hbar c \Delta_c}\right]^2 n_{\rm h}
\\ &\times\left[\left(\frac{m_{v}g_{\nu,v}(\Delta K)}{\hbar^2\Delta K^2}\right)^2 + \left(\frac{m_{c'}g_{\nu,c'}(\Delta K)}{\hbar^2\Delta K^2}\right)^2\right],
\end{split}
\end{equation}
 \begin{equation}\label{eq:G_d0x_phonon_gtr}
 \begin{split}
 &\Gamma^{>,\nu}_{\rm D^0X} \approx \frac{48\tilde{E}_{\rm g}}{\hbar}\frac{e^2}{\hbar c}\left[\frac{\gamma't_{vv}}{\hbar c \Delta_v}
 -\frac{\gamma t_{cc}}{\hbar c \Delta_c}\right]^2
 \\
 &\times \left[ \left(\frac{m_{v}g_{\nu,v}(\Delta K)}{\hbar^2\Delta K^2}\right)^2 + \left(\frac{m_{c'}g_{\nu,c'}(\Delta K)}{\hbar^2\Delta K^2}\right)^2\right]
 \\
&\times\int d^2 r \left| \int d^2 r'\,\chi^*(\rr')\Phi(\rr,\rr,\rr')\right|^2.
 \end{split}
 \end{equation}
 \end{subequations}
The VMC estimate of the overlap of $\chi({\bf r}')$ with $\Phi({\bf
  r},{\bf r},{\bf r}')$ is $3.85 \times 10^{-4}$ {\AA}$^{-2}$; see Table
\ref{table:overlaps}.

In the small-twist-angle limit $(<)$, phonon emission from D$^0_{c'}$h$_v$
complexes is dominated by the diagram of
Fig.\ \ref{fig:d0h_diagrams_phonons}(a). In that process, the phonon
is emitted by a hole in the WSe$_2$ layer, which then tunnels to
recombine with the electron bound to the donor impurity. By contrast,
all other diagrams shown in Fig.\ \ref{fig:d0h_diagrams_phonons}
involve ionization of the donor atom, which is suppressed by the large
binding energy of the D$^0_{c'}$ complex. The radiative rates for
D$^0_{c'}$h$_{v}$ can thus be approximated by (Appendix
\ref{app:phonon_assisted})
\begin{subequations}
\begin{equation}\label{eq:G_d0h_phonon_opt}
\begin{split}
&\Gamma^{<,\nu=\text{LO}/\text{HP}}_{\rm D^0h} \approx \frac{2\tilde{E}_{\rm g}}{\pi \hbar}\frac{e^2}{\hbar c}\left[\frac{\gamma't_{vv}}{\hbar c\Delta_v}\right]^2\left[ \frac{3\abs{g_{\nu,c'}(\Delta K)}^2}{\left(\hbar\omega_{\nu} +{\cal E}_{\rm D^0}^{\rm b} \right)^2}
\right.
\\
&+ \left. \frac{m_{v}|g_{\nu,v}(\Delta K)|^2\abs{F(\rr_0)}^2}{\hbar^3 \omega_{\nu}}\left|\int d^2 r\, e^{i\Delta \KK\cdot \rr}\chi(\rr)\right|^2 \right]n_{\rm h},
\end{split}
\end{equation}
\begin{equation}\label{eq:G_d0h_phonon_ac}
\begin{split}
&\Gamma^{<,\text{LA}}_{\rm D^0h} \approx \frac{2\tilde{E}_{\rm g}}{\pi  \hbar}\frac{e^2}{\hbar c}
\frac{m_v\,\Xi_{v}^2}{\hbar^2  \rho c_{\text{LA}}^2 } \left[\frac{\gamma't_{vv}}{\hbar c\Delta_v}\right]^2
\\
&\times  \left|\int d^2 r\, e^{i\Delta \KK\cdot \rr}\chi(\rr)\right|^2\abs{F(\rr_0)}^2n_{\rm h}.
\end{split}
\end{equation}
\end{subequations}
In the D${}_{c'}^0$X${}_{vc'}$ case at small twist angles, the phonon
emission process is suppressed by the ionization of the complex in the
intermediate state and the overlap integral between the initial
D${}_{c'}^0$X${}_{vc'}$ and final D${}_{c'}^0$ states. The rates are
given by
\begin{subequations}
\begin{equation}\label{eq:G_d0x_phonon_opt}
\begin{split}
&\Gamma_{\rm D^0X}^{<,\nu=\text{LO}/\text{HP}}=\frac{4\tilde{E}_{\rm g}}{\hbar}\frac{e^2}{\hbar c}\frac{\abs{F(\rr_0)}^2|g_{\nu,v}(0)|^2+3|g_{\nu,c'}(0)|^2}{(\hbar\omega_{\nu}+{\cal E}^{\rm b}_{\rm D^0X}+{\cal E}^{\rm b}_{\rm X})^2}
\\
&\times \left[\frac{\gamma't_{vv}}{\hbar c \Delta_v}
-\frac{\gamma t_{cc}}{\hbar c \Delta_c}\right]^2\int d^2 r \left| \int d^2 r'\,\chi^*(\rr')\Phi(\rr,\rr,\rr')\right|^2,
\end{split}
\end{equation}
\begin{equation}\label{eq:G_d0x_phonon_ac}
\begin{split}
  &\Gamma_{\rm D^0X}^{<,\text{LA}}=\frac{\tilde{E}_{\rm g}}{\sqrt{2}\hbar^3c_{\text{LA}}}\frac{e^2}{\hbar c}\frac{\left(m_v+m_{c'} \right)^{3/2}}{\sqrt{{\cal E}_{\rm D^0X}^{\rm b}+{\cal E}_{\rm X}^{\rm b}}}\left[\frac{\gamma't_{vv}}{\hbar c \Delta_v}
  -\frac{\gamma t_{cc}}{\hbar c \Delta_c}\right]^2
  \\
  &\times\left[ \frac{\abs{F(\rr_0)}^2\Xi_{v}^2}{\rho} \abs{\int d^2 r \int d^2r' \,\exp{-i\Delta\KK\cdot\rr}\chi^*(\rr')\Phi(\rr,\rr,\rr')}^2 \right.\\
  &\quad+\left. \frac{3\Xi_{c'}^2}{\rho'} \abs{\int d^2 r \int d^2r' \,\exp{i\Delta\KK\cdot\rr}\chi^*(\rr')\Phi(\rr,\rr,\rr')}^2 \right],
\end{split}
\end{equation}
\end{subequations}
where $\rho$ and $\rho'$ are the mass densities of WSe$_2$ and MoSe$_2$,
respectively (Table \ref{tab:phonon_params}).

In Eqs.\ (\ref{eq:G_d0h_phonon_gtr})
--(\ref{eq:G_d0x_phonon_ac}), the electron-layer contributions to the decay rate contain a factor of three originating from the tunneling process, which gives three distinct intermediate states with different emitted
phonon wave vectors, related by $C_3$ symmetry. As a result, the
interference factor appearing in the interaction-driven
processes of Secs.\ \ref{sec:complexes} and \ref{sec:asymptotics} is
absent in this case. For the hole layer, however, the interference factor remains due to the momentum spread of the complex wave function, which lifts the requirement that the hole be scattered exactly onto the electron-layer valley in order to recombine.

Additional contributions to the LO phonon emission come from e-ph interaction of a carrier in one layer with an LO
phonon in the other. This is made possible by the long range of the LO
phonon-induced potential. The interlayer separation results
in an exponential suppression of the potential in the interlayer
distance and momentum transfer as $e^{-\Delta K d}$, which nonetheless
is approximately unity in the limit of close alignment. Thus, we add this contribution to the LO-phonon-assisted recombination rates for D${}_{c'}^0$h${}_v$ and D${}_{c'}^0$X${}_{vc'}$ complexes in the small-twist-angle limit.

The total phonon emission rates for the two complexes, combining the
three phonon modes, are shown in Fig.\ \ref{fig:figure1}(b) as
functions of the twist angle.
As mentioned above, the phonon contribution to the recombination rate is
most significant for the D$^0_{c'}$h$_v$ complex, being an order of magnitude
larger than for D$^0_{c'}$X$_{vc'}$. The LO phonon mode in the
hole layer (WSe$_2$) is the dominant phonon-assisted process overall,
and gives a significant decay rate in the
small-twist-angle limit. As a result, we predict additional phonon-replica lines
in the PL spectrum, red shifted by the phonon energy
$\hbar\omega_{\text{LO}}=31$ meV with respect to the main
D$_{c'}^0$h$_v$ and D$_{c'}^0X_{vc'}$ lines. The D$_{c'}^0$h$_v$ phonon-replica line gives the most dominant feature, with decay rates comparable to the main D$_{c'}^0$h$_v$ line.

\begin{figure}[!t]
  \centering
  \includegraphics[width=0.95\columnwidth]{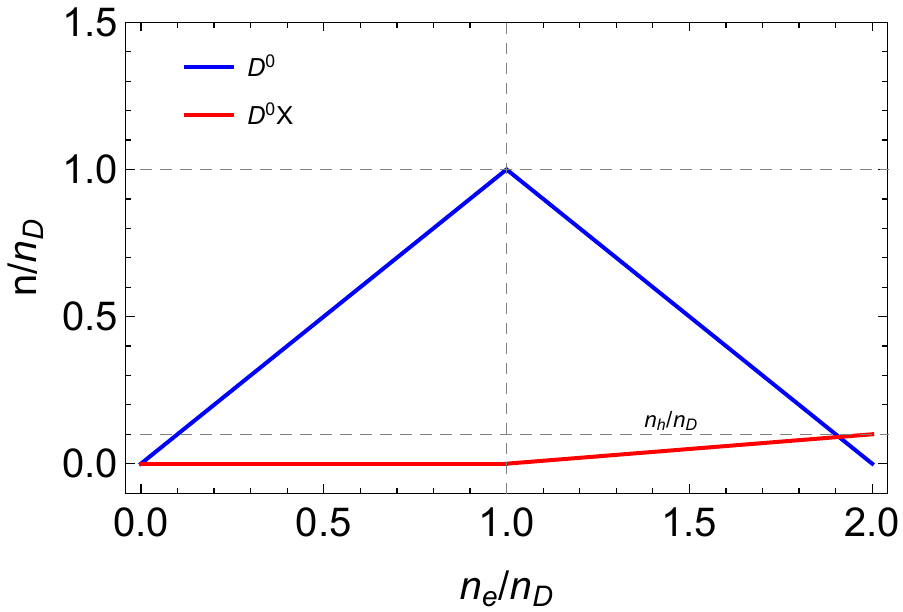}
  \caption{Model for the density of complexes D$^0_{c'}$ and D$^0_{c'}$X$_{vc'}$ as a
    function of the electron density $n_{\rm e}$.}
  \label{fig:occ}
\end{figure}

\section{Intensity dependence on doping}
\label{sec:intens}

In addition to the decay rates, the relative line intensities also
depend on the distribution of $\mathrm{D}_{c'}^0\mathrm{h}$ and
$\mathrm{D}_{c'}^0\mathrm{X}_{vc'}$ complexes in the system. At charge
neutrality, neutral excitonic complexes such as
$\mathrm{D}_{c'}^0\mathrm{h}_v$ are energetically favorable, whereas
additional electrons introduced into the sample will bind to existing
neutral donors to form $\mathrm{D}_{c'}^0\mathrm{X}_{vc'}$
complexes. Thus the relative population of complexes can be
controlled through doping.

In this section we model the evolution of the PL spectrum with the electron
carrier density within the range $0< n_{\rm e}\le 2n_{\rm D}$, controlled 
by means of gating\cite{vialla_private}. We use a simplified zero-temperature model for the occupations of the
two complexes, shown in Fig.\ \ref{fig:occ}. There are 
two main regimes determined by the sample-dependent donor density $n_{\rm D}$.  In the p-doped regime, defined by $0<n_{\rm e}<n_{\rm D}$, added electrons
neutralize the excess positive donors, forming $\mathrm{D}_{c'}^0$
complexes that can recombine with the optically pumped holes. In this
regime, the formation of $\mathrm{D}_{c'}^0\mathrm{X}_{vc'}$ complexes
is energetically unfavorable, and thus thermally suppressed until all
donors have been neutralized. By contrast, in the n-doped regime, defined
by $n_{\rm D}<n_{\rm e}<2n_{\rm D}$, it is energetically favorable for additional
electrons to bind with an existing neutral donor to form either a charged donor state $\mathrm{D}_{c'c'}^-$ (Table \ref{tab:dmc_total_energy}), or a donor-bound trion $\mathrm{D}_{c'}^0\mathrm{X}_{vc'}$. For the latter case we must consider that laser-pumped holes are scarce ($n_{\rm h} \ll n_{\rm D}$), and thus the probability of forming a $\mathrm{D}_{c'}^0\mathrm{X}_{vc'}$ complex will be proportional to $n_{\rm h}/n_{\rm D}$. The increase in electron density is accompanied by a
decrease in $\mathrm{D}_{c'}^0\mathrm{h}_{v'}$ numbers, and a much slower increase in the $\mathrm{D}_{c'}^0\mathrm{X}_{vc'}$ population, until the number of donor-bound trions in the system equals the number of
available holes. This is shown in Fig.\ \ref{fig:occ}, and can be summarized as
\begin{equation}
n_{\rm D^0} =
\begin{cases}
n_{\rm e},& n_{\rm e}<n_{\rm D}
\\
n_{\rm D}\left[1-\frac{n_{\rm e}-n_{\rm D}}{n_{\rm D}}\right], &  n_{\rm D}<n_{\rm e}<2n_{\rm D}
\end{cases},
\label{eq:int1}
\end{equation}
and
\begin{equation}
n_{\rm D^0X} =
\begin{cases}
0, & n_{\rm e}<n_{\rm D}
\\
n_{\rm D}\frac{n_{\rm h}}{n_{\rm D}}\frac{n_{\rm e}-n_{\rm D}}{n_{\rm D}},& n_{\rm D}<n_{\rm e}<2n_{\rm D}
\end{cases}.
\label{eq:int2}
\end{equation}

Eqs.\ \eqref{eq:int1} and \eqref{eq:int2}, together with \eqref{eq:d0h_lesser}, show the dependence
of $n_{\rm D^0}$ and $n_{\rm D^0X}$ on the hole density. This dependence is critical for radiative
recombination, given the scarcity of holes by comparison to the donor density. Thus, to give a realistic
estimate of the intensity, we consider the effects of non-radiative recombination of holes through impurity-driven
processes. The density of holes lost through these processes per unit time can be written as $\tau_{0}^{-1}n_{\rm h}$,
where $\tau_0^{-1}$ is the non-radiative decay rate. Assuming that holes are laser-pumped at a constant rate $\tau_{\rm pump}^{-1}n_0$, where
$n_0$ is a constant with dimensions of inverse area, the hole density obeys
the rate equation
\begin{equation}
	\dot{n}_{\rm h} = \tau_{\rm pump}^{-1}n_{0} - \tau_0^{-1}n_{\rm h},
\end{equation}
with the steady state solution $n_{\rm h} = \tau_0\tau_{\rm pump}^{-1}n_0$. In the p-doped regime,
delocalized holes can recombine non-radiatively with the electrons present in the sample, and the
non-radiative lifetime can be assumed of the form $\tau_0 = c_0/n_{\rm e}$, with $c_0$ a constant. Thus, 
writing the D${}_{c'}^0$h${}_{v}$ radiative intensity  as $I_{\rm D^0h}=\Gamma_{\rm D^0h}n_{\rm D^0}$,
we obtain the expression
\begin{equation}\label{eq:ID0h}
\begin{split}
	I_{\rm D^0h}=&\frac{4 \tilde{E}_{\rm g}\abs{F(\rr_0)}^2}{\hbar}\frac{e^2}{\hbar c}\left[
\frac{t_{vv}\gamma'}{\hbar c\Delta_v}-\frac{t_{cc}\gamma}{\hbar c(\Delta_c+{\cal E}^{\rm b}_{\rm D^0})}
\right]^2
\\
&{} \times \left|
\int d^2r\, e^{i\Delta \KK\cdot \rr }\chi(\rr)\right|^2c_1,
\end{split}
\end{equation}
where $c_1 =  c_0n_0\tau_{\rm pump}^{-1}$ is a constant independent of the electron density.

A similar argument can be made for the n-doped regime. In this case, the intensity is given by
$I_{\rm D^0X}=\Gamma_{\rm D^0X}n_{\rm D^0X}$, where the number of donor-bound trions
can be approximated as $n_{\rm D^0X}=n_{\rm h}(n_{\rm e} - n_{\rm D})/n_{\rm D}$. However,
in this regime the holes will be localized near the donor-impurity sites forming D${}_{c'}^0$X${}_{vc'}$
states, where they will be in close proximity to two electrons with which they can recombine
non-radiatively. Thus, we may approximate the non-radiative decay rate as
$\tau_0 = c_0/2n_{\rm D}$. This leads to
\begin{equation}\label{eq:ID0X}
	I_{\rm D^0X} = \frac{(n_{\rm e} - n_{\rm D})}{2n_{\rm D}^2}c_1 \Gamma_{\rm D^0X}.
\end{equation}

\begin{figure}[t!]
  \centering
  \includegraphics[width=1\columnwidth]{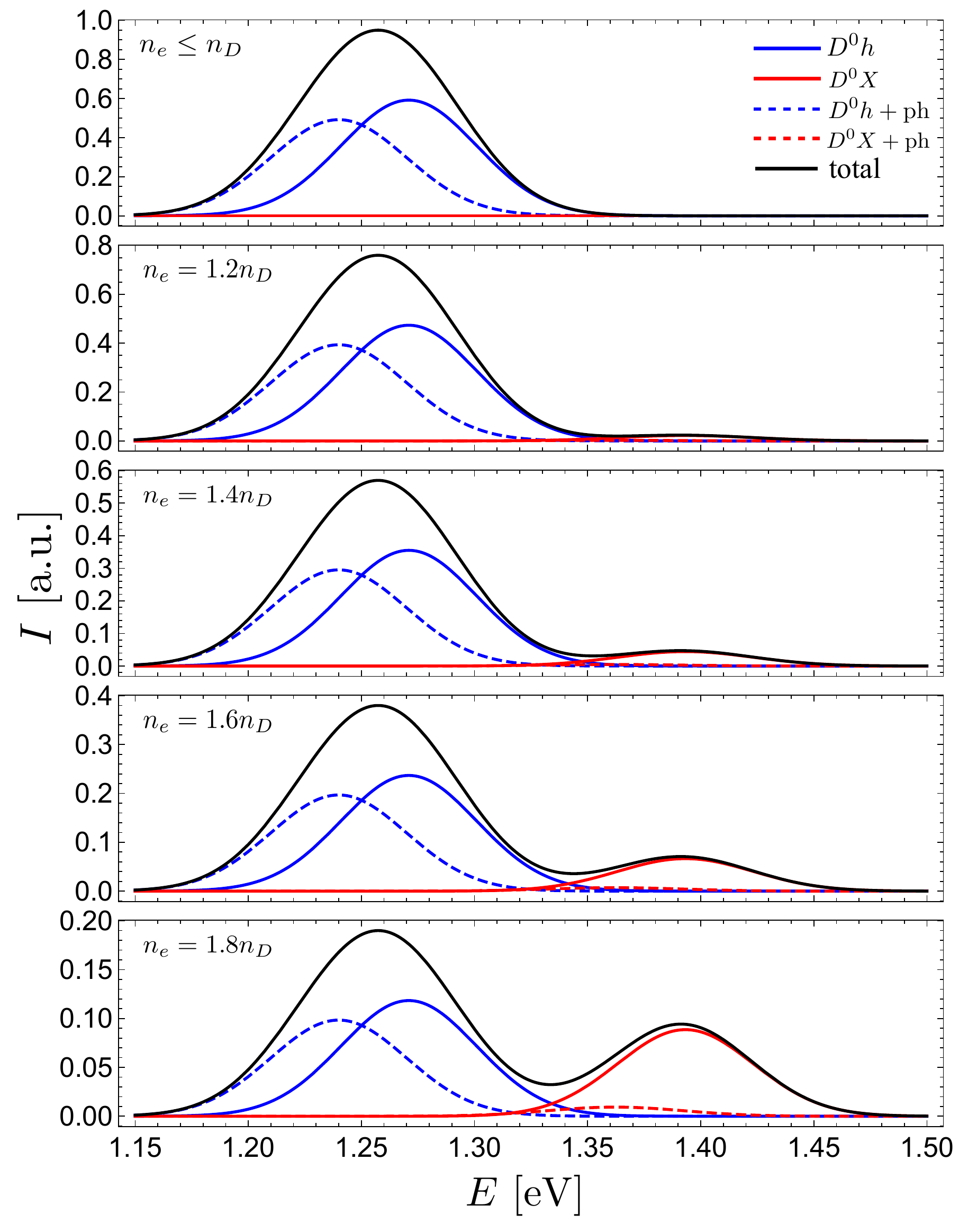}
  \caption{Simulated normalized PL spectra for a closely aligned ($\theta \approx 0^\circ$) MoSe${}_2$/WSe${}_2$ heterobilayer, originating from the
    D$^0_{c'}$h$_{v}$ and D$^0_{c'}$X$_{vc'}$ complexes at different
    electron densities $n_{\rm e}$, given in terms of the fixed donor density
    $n_{\rm D}$. Dashed curves correspond to the phonon replicas. The lines are assumed to have Gaussian shapes of width $2\sigma=60$ meV, and we use $n_{\rm h}=10^{11}$ cm$^{-2}$ and $n_{\rm D}=10^{13}$ cm$^{-2}$.}
  \label{fig:spectrum}
\end{figure}
The resulting simulated PL spectrum is shown in
Fig.\ \ref{fig:spectrum} for different doping densities, given in
terms of the donor density in the MoSe$_2$ layer. A Gaussian line-shape was used for the lines with an experimentally motivated broadening \cite{vialla_private} of $2\sigma=60$ meV\@.
The spectrum shows the three dominant lines, D${}_{c'}^0$h$_v$, D${}_{c'}^0$X${}_{vc'}$, and
the red-shifted phonon replica of D$^0_{c'}$h$_v$, with the lines' peak
energies determined by the DMC-obtained binding energies. The three
complexes evolve with doping as prescribed by Eqs.\ \eqref{eq:ID0h} and \eqref{eq:ID0X}.
The D${}_{c'}^0$h$_v$ complex
and its phonon replica dominate for $0<n_{\rm e}<n_{\rm D}$; then, the
D${}_{c'}^0$X${}_{vc'}$ line grows slowly in intensity in the n-doped regime, with a simultaneous reduction in the
intensity of the D${}_{c'}^0$h$_v$ complex.  For the broadening used in
the simulated PL spectrum, the proximity of the three lines
results in an intricate line form, providing a signature in
PL experiments for the intrinsic structure of the interlayer emission line.

\section{Conclusions}\label{sec:conclusions}
The momentum mismatch between twisted and incommensurate heterobilayer
TMDs prevents efficient radiative recombination of interlayer
complexes composed of electrons and holes localized on opposite
layers. In this paper we described mechanisms that  bridge the
momentum gap involving donor impurities present in the heterobilayer
system, both at small and large twist angles. The donor impurities
were found to provide deep potential wells ($\sim 200$ meV), resulting
in strongly bound interlayer complexes, as revealed by DMC
calculations. Focusing on the simplest multiparticle complexes, we
estimate radiative rates of up to a few $\mu$s$^{-1}$ for the neutral
donor with a free hole D$^0_{c'}$h$_{v}$ and the donor-bound trion
D$^0_{c'}$X$_{vc'}$ complexes for closely aligned layers, and a strong twist-angle
suppression for large misalignment with the asymptotic form $
\propto\theta^{-8}$. A comparable contribution was found for the D$^0_{c'}$h$_{v}$
complex from emission of optical phonons, resulting in a total of
three dominant and doping-tunable lines in the PL spectrum. The
D$^0_{c'}$h$_{v}$ line
and its phonon replica are expected to dominate the emission spectrum
for electron densities below the sample-dependent donor concentration; conversely,
PL from the D$^0_{c'}$X$_{vc'}$ complex is expected to dominate the interlayer
sector of the spectrum when the electron density exceeds the density of donors.

Based on QMC simulations, we have
shown that our qualitative results are robust against uncertainty in
model parameters, such as the band effective masses, as well as
sample-dependent dielectric properties. Therefore, our predictions
provide a new perspective for interpreting recent experimental
observations of interlayer luminescence in heterobilayers of
transition-metal dichalcogenides.

All relevant  data  present  in  this  publication  can  be  accessed  at Lancaster University\footnote{\href{https://dx.doi.org/10.17635/lancaster/researchdata/219}{https://dx.doi.org/10.17635/lancaster/researchdata/219}}.

\acknowledgments{M.D., D.A.R-T., and V.I.F.\ wish to thank F.\ Vialla and F.\ Koppens for fruitful discussions.
M.D., D.A.R-T., and V.I.F.\ acknowledge support from
ERC Synergy Grant Hetero2D, EPSRC EP/N010345, EPSRC EP/P026850/1,
Lloyd Register Foundation Nanotechnology grant, and support from the European Graphene Flagship Project.
 R.J.H.\ is fully funded by the Graphene NOWNANO center for doctoral training (EPSRC grant no.\ EP/L01548X/1).
  M.S.\ was funded by the EPSRC standard grant
``Non-perturbative and stochastic approaches to many-body localization'' (Grant No.\ EP/P010180/1).
Computer time was provided by Lancaster University's High-End Computing facility. This work made use of the facilities
of N8 HPC Centre of Excellence, provided and funded by the N8 consortium and EPSRC (Grant No.EP/K000225/1). The Centre
is co-ordinated by the Universities of Leeds and Manchester.

}

\bibliographystyle{apsrev4-1}
\bibliography{refs.bib}

\appendix

\section{Long-range interaction between charge carriers}
\label{sec:multilayer_keldysh}

\subsection{Multilayer Keldysh interaction}

Consider a vdW heterostructure of 2D semiconductors comprised of $N$
parallel layers (labelled $i=1,2,\ldots,N$), each having in-plane
susceptibility $\kappa_i$ and $z$-coordinate $d_i$.  Suppose this
heterostructure is immersed in an isotropic medium of dielectric
constant $\epsilon$.  In practice the dielectric constant is taken to
be the average of the dielectric constants of the media above and
below the heterobilayer.

Suppose that a test charge density
\begin{equation}
  \rho^j_{\text{tot}}({\bf r}, z)= \rho^j({\bf r}) \delta(z - d_j),
\end{equation}
is present in layer $j$.
The resulting electric displacement field is
\begin{equation}
  {\bf D} = -\frac{\epsilon}{4\pi}\nabla\phi({\bf r}, z) -
  \sum_i\kappa_i [\nabla_{\parallel}\phi({\bf r}, d_i)]\delta(z-d_i),
\end{equation}
where $\nabla_{\parallel}$ is the 2D gradient operator (excluding the
$z$-component). Gauss's law yields
\begin{eqnarray}
  \rho^j({\bf r}, z) \delta(z-d_j) & = &
  -\frac{\epsilon}{4\pi}\nabla^2\phi({\bf r},z) \nonumber \\ & & {} -
  \sum_i \kappa_i[\nabla^2_{\parallel}\phi({\bf
      r},d_i)]\delta(z-d_i). \nonumber \\
\end{eqnarray}
Taking the Fourier transform gives
\begin{equation}
  \rho^j({\bf q})\exp{-ikd_j} =
  \frac{\epsilon}{4\pi}(q^2+k^2)\phi({\bf q},k) + q^2 \sum_i \kappa_i
  \phi({\bf q}, d_i) \exp{-ikd_i},
\end{equation}
which, after Fourier inversion in the $k$ variable only, gives
\begin{equation}
  \rho^j({\bf q})\exp{-q\lvert z - d_j \rvert} = \frac{\epsilon}{2\pi}
  q \phi({\bf q}, z) + q^2 \sum_i \kappa_i \phi({\bf q},
  d_i)\exp{-q\lvert z - d_i \rvert}.\label{eq:rhoj_z}
\end{equation}
Evaluating Eq.\ (\ref{eq:rhoj_z}) at each layer ($z=d_l$, $l=
\{1,2,\ldots,N\}$), we find
\begin{eqnarray}
  \rho^j({\bf q})\exp{-q\lvert d_l -d_j \rvert} & = & q[\epsilon/(2\pi) -
  \kappa_l q] \phi({\bf q}, d_l) \nonumber \\ & & {} + q^2 \sum_{i \neq l} \kappa_i
  \phi({\bf q}, d_i)\exp{-q\lvert d_l - d_i \rvert},
\end{eqnarray}
which is a matrix equation
\begin{equation}
  \rho^j_{l}({\bf q}) = \sum_i M_{l i}({\bf q}) \phi_{i}({\bf q}),
  \label{eq:mat_eqn}
\end{equation}
where
\begin{align}
  \rho^j_{l}({\bf q}) &= \rho^j({\bf q})\exp{-q \lvert d_l - d_j
    \rvert}, \nonumber \\ \phi_{i}({\bf q}) &= \phi({\bf q}, d_{i}),
  \nonumber \\ M_{l i} &= \left\{ \begin{array}{ll} q[\epsilon/(2\pi) +
    \kappa_{l} q] & \mbox{if~}i=l \\ q^2 \kappa_{i} \exp{-q \lvert
      d_{l} - d_{i} \rvert} & \mbox{otherwise} \end{array}
  \right. .\label{eq:definitions}
\end{align}

The solution to Eq.\ (\ref{eq:mat_eqn}) is a set of $\phi_i({\bf
  q}) \equiv \rho^j({\bf q}) \times v_{ji}({\bf q})$, with $v_{ji}({\bf q})$
being the Fourier components of the interaction potential between
layer $j$ and layer $i$. If $j=i$ then this is the intralayer
interaction in layer $j$. This procedure should, in general, be
repeated for $j = 1,2,\ldots,N$; however, if there is sufficient
symmetry (e.g., a mirror symmetry about a plane through the center of
the heterostructure) then only a subset of $j$ values will require
explicit solution of Eq.\ (\ref{eq:mat_eqn}).

The same analysis can be shown to apply in the case that the
surrounding dielectric medium is anisotropic, having dielectric tensor
\begin{equation}
  \tilde \epsilon = \begin{pmatrix}
  \epsilon_{\parallel} & 0 & 0 \\
  0 & \epsilon_{\parallel} & 0 \\
  0 & 0 & \epsilon_{\perp}
  \end{pmatrix},
\end{equation}
provided the substitutions
\begin{eqnarray}
  d_i \rightarrow D_i &=& \sqrt{\epsilon_{\parallel}/\epsilon_{\perp}} d_i, \\
  \epsilon \rightarrow \bar{\epsilon} &=& \sqrt{\epsilon_{\parallel}
  \epsilon_{\perp}},
\end{eqnarray}
are also made.


\subsection{Numerical evaluation of the bilayer Keldysh interaction}
\label{sub:bilayer_keldysh}

In the bilayer case ($N=2$), it is straightforward to solve
Eq.\ (\ref{eq:mat_eqn}) to obtain the intralayer ($\mathcal{V}$ and
$\mathcal{V}'$) and interlayer ($\mathcal{W}$) potentials of
Eqs.\ (\ref{eq:exact_int_11})--(\ref{eq:exact_int_12}).

Continuum QMC calculations require the potential energy to be
evaluated in real space. We therefore require the inverse Fourier
transforms of Eqs.\ (\ref{eq:exact_int_11})--(\ref{eq:exact_int_12}),
which reduce to Hankel transforms due to the circular symmetry of the
interaction potentials.

At long range (small $q$), the intralayer interaction $\mathcal{V}({\bf
  q})=2\pi/\left\{\epsilon q[1+(r_{*}+r_{*}')q]\right\}+O(q)$ reduces
to the monolayer Keldysh form \cite{keldysh}, with an effective
screening length $r_{*}^{\text{eff}}=r_{*}+r_{*}'$. The inverse Fourier transform
can be performed analytically in this limit, giving
\begin{eqnarray} \mathcal{V}(r) & \approx & \frac{\pi}{2\epsilon (r_{*}+r_{*}')} \nonumber \\ & & {} \times \left[ H_0(r/(r_{*}+r_{*}'))-Y_0(r/(r_{*}+r_{*}')) \right] \nonumber \\ & & {} + O(r^{-3}), \label{eq:approx_intralayer} \end{eqnarray}
where $H_0$ and $Y_0$ are a Struve function and a Bessel function of
the second kind, respectively.  Equation (\ref{eq:approx_intralayer})
is a good approximation at long range.

At short range (large $q$), the intralayer interaction of
Eq.\ (\ref{eq:exact_int_11}) again reduces to the monolayer Keldysh
form, but this time with $r_{*}^{\text{eff}}=r_{*}$, i.e., the second layer becomes
irrelevant.  On the other hand, at very
long range, the monolayer Keldysh interaction is also valid, since
$\mathcal{V}({\bf q})=2\pi/(\epsilon q)+O(1)$ at small $q$ so that the
interaction is of Coulomb form.  Thus the monolayer Keldysh
interaction
\begin{equation} \mathcal{V}(r) \approx \frac{\pi}{2\epsilon r_{*}^{\text{eff}}} \left[ H_0(r/r_{*}^{\text{eff}})-Y_0(r/r_{*}^{\text{eff}}) \right] + O(r^{-2}), \label{eq:approx_mono_intralayer} \end{equation}
is a reasonable approximation to the intralayer interaction at
\textit{both} short and very long range.

To evaluate the ``full'' intralayer interaction numerically, we used
the quadrature method of Ogata \cite{ogata2005numerical} to perform
the Hankel transform of $\mathcal{V}({\bf q})-2\pi/\left\{ \epsilon
q[1+r_{*}q]\right\}$, then added the result to the monolayer Keldysh
interaction of Eq.\ (\ref{eq:approx_mono_intralayer}).  Partitioning the
interaction into a long-range part and a numerically evaluated
short-range part ensures that the quadrature is relatively
straightforward, and that we can introduce a cutoff at large $r$,
beyond which the numerical corrective term is negligible.

At small $q$, the interlayer interaction of
Eq.\ (\ref{eq:exact_int_12}) reduces to the displaced
Coulomb form $\mathcal{W}(q)=2\pi\exp{-(r_{*}+r_{*}'+d)q}/(\epsilon q) + O(q)$;
hence the long-range interlayer potential in real space is given by
\begin{equation}
  \mathcal{W}(r) \approx \dfrac{1}{4\pi\sqrt{r^2+{(r_{*}+r_{*}'+d)}^2}} +
  O(r^{-3}). \label{eq:coul_approx_interlayer}
\end{equation}
At short range in real space the interlayer interaction should be
nondivergent.  Equation (\ref{eq:coul_approx_interlayer})
satisfies this qualitative requirement.

To evaluate the ``full'' interlayer interaction numerically, we
performed the numerical Hankel transform of $\mathcal{W}({\bf
  q})-2\pi\exp{-(r_{*}+r_{*}'+d)q}/(\epsilon q)$, then added the
result to Eq.\ (\ref{eq:coul_approx_interlayer}).

There is an alternative long-range approximation to the interlayer
potential, which is more like the intralayer potential.  Noting
that $\mathcal{W}({\bf q})=2\pi/\left\{\epsilon
  q[1+(r_{*}+r_{*}'+d)q]\right\}+O(q)$, the long-range interlayer potential
reduces to a Keldysh potential with $r_*^{\text{eff}}=r_{*}+r_{*}'+d$, giving
\begin{eqnarray} \mathcal{W}(r) & \approx & \frac{\pi}{2\epsilon (r_{*}+r_{*}'+d)} \nonumber \\ & & {} \times \left[ H_0\left(\frac{r}{r_{*}+r_{*}'+d}\right)-Y_0\left(\frac{r}{r_{*}+r_{*}'+d}\right) \right] \nonumber \\ & & {} + O(r^{-3}). \label{eq:approx_interlayer} \end{eqnarray}
This introduces unphysical singular behavior into the interlayer
interaction at short range.

\section{QMC calculations}
\label{sec:mc}

\subsection{Technical details} \label{sec:technical}

We performed VMC and DMC calculations
\cite{Ceperley_1980,Foulkes_2001} for complexes of distinguishable
charge carriers and fixed ions interacting via the ``full'' bilayer
potential
[Eqs.\ (\ref{eq:exact_int_11})--(\ref{eq:exact_int_12})] and the
approximate small-$q$ Keldysh form of the potential
[Eqs.\ (\ref{eq:approx_intralayer}) and (\ref{eq:approx_interlayer})],
as described in Appendix \ref{sub:bilayer_keldysh}.  We used trial
wave functions of Jastrow form, where the Jastrow exponents contained
smoothly truncated polynomial particle-particle terms, ion-particle
terms, ion-particle-particle, and particle-particle-particle terms
\cite{Drummond_2004,Lopez_2012}. Additional terms satisfying the
analogs of the Kato cusp conditions
\cite{ganchev,marcin_bindingenergies_prb_2017,kato1957eigenfunctions}
were applied to the trial wave function between pairs of particles
wherever there was a logarithmic divergence in the interaction between
them, including the unphysical divergences in the approximate Keldysh
interaction.  Free parameters were optimized using VMC with variance
\cite{Umrigar_1988,Drummond_2005} and energy minimization
\cite{Umrigar_2007} as implemented in the \textsc{casino} code
\cite{needs2009continuum}.

In our DMC calculations we used two DMC time steps in the ratio $1 {:}
4$ and corresponding target populations in the ratio $4 {:} 1$,
allowing a simultaneous extrapolation to zero time step and infinite
population.  Since the charge carriers are distinguishable, there is
no fixed-node error and hence DMC provides exact ground-state
solutions to the effective-mass model of interacting charge carriers
with the chosen model interaction.

\subsection{Energies of complexes in the hBN/MoSe$_2$/WSe$_2$/hBN
  heterostructure}
\label{sub:dmc_energies}

Table \ref{tab:dmc_total_energy} shows the total energies of charge-carrier
complexes in the hBN/MoSe$_2$/WSe$_2$/hBN heterostructure. For
completeness we include results in which the electrons are found in
either layer; however, the results of immediate relevance to this
paper are those for which the electrons are all found in the MoSe$_2$
layer.  DMC results for two-particle complexes agree
with calculations performed using Mathematica's finite-element method
\cite{Mathematica} (see Appendix \ref{sec:finite}). Using total
energies, one can assess the most energetically favorable
dissociations (see Table \ref{tab:dmc_dissociations}) and therefore
calculate the binding energies of the various complexes.

\begin{table}[tb]
  \caption{DMC total energies of various charge-carrier complexes in
    the hBN/MoSe$_2$/WSe$_2$/hBN heterostructure calculated using the
    monolayer Keldysh approximation to the bilayer potential
    [Eqs.\ (\ref{eq:approx_intralayer}) and
      (\ref{eq:approx_interlayer})] and using the full bilayer
    interaction
    [Eqs.\ (\ref{eq:exact_int_11})--(\ref{eq:exact_int_12})].  Primes
    ($'$) indicate that a charge carrier is in the MoSe$_2$ layer;
    otherwise the charge carrier is in the WSe$_2$ layer. The
    subscripts $c$ and $v$ indicate whether charge carriers are
    electrons ($c$) or holes ($v$).  Donor ions are always assumed to
    be in the MoSe$_2$ layer, while acceptor ions are always assumed
    to be in the WSe$_2$ layer. Interlayer complexes in which all the
    electrons are in the MoSe$_2$ layer and all the holes are in the
    WSe$_2$ layer are listed in the upper section of the table;
    complexes in which some of the electrons are in the WSe$_2$ layer
    are listed in the lower section of the table.
\label{tab:dmc_total_energy}}
  \begin{tabular}{lr@{.}lr@{.}l}
    \hline \hline

    & \multicolumn{4}{c}{DMC total energy (meV)} \\

    \raisebox{1.5ex}[0pt]{Complex} &
    \multicolumn{2}{c}{Approx.\ Keldysh} & \multicolumn{2}{c}{Bilayer
      potential} \\

    \hline

    X$_{vc'}$                & ~~$-103$&$958669(5)$  & ~~~~$-84$&$232(1)$\\

    X$^-_{vc'c'}$            & $-108$&$1967(4)$      & $-88$&$32(3)$\\

    X$^+_{vvc'}$             & \multicolumn{2}{c}{ } & $-88$&$12(2)$ \\

    X$_{vc'}$X$_{vc'}$            & \multicolumn{2}{c}{ } & \multicolumn{2}{c}{unbound}\\

    X$_{vc'}$X$^-_{vc'c'}$        & \multicolumn{2}{c}{ } & \multicolumn{2}{c}{unbound}\\

    D$^0_{c'}$               & $-163$&$2478711(5)$   & $-229$&$03306(1)$\\

    D$^-_{c'c'}$             & $-176$&$9426(3)$      & $-249$&$60(2)$ \\

    D$^0_{c'}$h$_{v}$        & $-163$&$4819(8)$      & \multicolumn{2}{c}{unbound}\\

    D$^0_{c'}$X$_{vc'}$      & $-278$&$73(2)$        & $-335$&$781(4)$\\

    D$^0_{c'}$X$^+_{vvc'}$   & \multicolumn{2}{c}{ } & $-340$&$891(6)$ \\

    D$^-_{c'c'}$X$_{vc'}$    & $-292$&$83(1)$        & $-343$&$26(3)$\\

    D$^0_{c'}$X$_{vc'}$X$_{vc'}$  & \multicolumn{2}{c}{ } & \multicolumn{2}{c}{unbound}\\

    D$^-_{c'c'}$X$_{vc'}$X$_{vc'}$& \multicolumn{2}{c}{ } & $-430$&$9(1)$\\

    A$^{0}_{v}$              & \multicolumn{2}{c}{ } & $-205$&$24083(1)$ \\

    A$^{+}_{vv}$             & \multicolumn{2}{c}{ } & $-223$&$56(1)$ \\

    A$^0_v$e$_{c'}$          & \multicolumn{2}{c}{ } & \multicolumn{2}{c}{unbound} \\

    A$^{0}_{v}$X$_{vc'}$     & \multicolumn{2}{c}{ } & $-309$&$411(4)$ \\

    A$^0_v$X$^-_{vc'c'}$     & \multicolumn{2}{c}{ } & $-315$&$021(8)$ \\

\hline

    X$_{vc}$                 & $-114$&$601814(1)$  & $-140$&$4303329(4)$ \\

    D$^0_{c}$                & $-124$&$890219(9)$  & $-102$&$5996(7)$ \\

    X$^-_{vcc'}$             & $-120$&$6018(5)$    & \multicolumn{2}{c}{unbound} \\

    X$^-_{vcc}$              & $-123$&$7189(5)$    & $-152$&$25(1)$ \\

    D$^-_{cc'}$              & $-165$&$8499(5)$    & \multicolumn{2}{c}{unbound} \\

    D$^-_{cc}$               & $-129$&$3199(9)$    & \multicolumn{2}{c}{unbound} \\

    D$^+$X$_{vc}$            & $-133$&$758(2)$     & $-141$&$716(8)$ \\

    D$^0_{c'}$X$_{vc}$       & $-279$&$776(5)$     & \multicolumn{2}{c}{unbound} \\

    D$^-_{c'c'}$X$_{vc}$     & $-301$&$81(1)$      & \multicolumn{2}{c}{unbound} \\

    D$^0_{c'}$X$^-_{vcc}$    & $-295$&$00(1)$      & \multicolumn{2}{c}{unbound} \\

    \hline \hline
  \end{tabular}
\end{table}

\begin{table}[tb]
  \caption{Dissociations of complexes and
    the associated binding energies in hBN/MoSe$_2$/WSe$_2$/hBN\@.
    The naming convention for the carrier complexes is explained in
    the caption of Table
    \ref{tab:dmc_total_energy}. \label{tab:dmc_dissociations}}
  \begin{tabular}{lllr@{.}lr@{.}l}
    \hline \hline

    &  &  & \multicolumn{4}{c}{Binding energy (meV)} \\

    \multicolumn{3}{l}{\raisebox{1.5ex}[0pt]{Dissociation process}} & \multicolumn{2}{c}{Appr.\ Kel.}
    & \multicolumn{2}{c}{Bilayer pot.}\\

    \hline

    X$^-_{vc'c'}$ & $\rightarrow$ & X$_{vc'}+{}$e$_{c'}$ &
      ~~$4$&$2380(4)$ & ~~~~$4$&$09(3)$\\

    X$^{+}_{vvc'}$ & $\rightarrow$ & X$_{vc'}+{}$h$_{v}$ & \multicolumn{2}{c}{ } & $3$&$89(2)$ \\

    X$_{vc'}$X$_{vc'}$ & $\rightarrow$ & X$_{vc'}+{}$X$_{vc'}$ & \multicolumn{2}{c}{ } & \multicolumn{2}{c}{unbound} \\

    X$_{vc'}$X$^{-}_{vc'c'}$ & $\rightarrow$ & X$_{vc'}+{}$X$^-_{vc'c'}$ & \multicolumn{2}{c}{ } & \multicolumn{2}{c}{unbound} \\

    D$^-_{c'c'}$ & $\rightarrow$ & D$^0_{c'}+{}$e$_{c'}$ &
    $13$&$6948(3)$ & $20$&$57(1)$ \\

  D$^0_{c'}$h$_{v}$ & $\rightarrow$ & D$^0_{c'}+{}$h$_{v}$ &
    $0$&$2340(8)$ & \multicolumn{2}{c}{unbound}\\

    D$^0_{c'}$X$_{vc'}$ & $\rightarrow$ & X$_{vc'}+{}$D$^0_{c'}$ & $11$&$52(2)$ & $22$&$516(4)$ \\

    D$^{0}_{c'}$X$^{+}_{vvc'}$ & $\rightarrow$ & D$^0_{c'}+{}$X$^{+}_{vvc'}$ & \multicolumn{2}{c}{ } & $23$&$74(2)$ \\

    D$^-_{c'c'}$X$_{vc'}$ & $\rightarrow$ & X$_{vc'}+{}$D$^-_{c'c'}$ & $11$&$93(1)$ & $9$&$43(4)$ \\

  D$^{0}_{c'}$X$_{vc'}$X$_{vc'}$ & $\rightarrow$ & D$^0_{c'}$X$_{vc'}+{}$X$_{vc'}$ & \multicolumn{2}{c}{ } & \multicolumn{2}{c}{unbound} \\

  D$^{-}_{c'c'}$X$_{vc'}$X$_{vc'}$ & $\rightarrow$ & D$^{-}_{c'c'}$X$_{vc'}+{}$X$_{vc'}$ & \multicolumn{2}{c}{ } & $3$&$3(2)$ \\

  A$^{+}_{vv}$ & $\rightarrow$ & A$^0_{v}+{}$h$_{v}$  & \multicolumn{2}{c}{ } & $18$&$32(1)$ \\

  A$^0_v$e$_{c'}$ & $\rightarrow$ & A$^0_{v}+{}$e$_{c'}$ & \multicolumn{2}{c}{ } & \multicolumn{2}{c}{unbound} \\

  A$^{0}_{v}$X$_{vc'}$ & $\rightarrow$ & A$^0_{v}+{}$X$_{vc'}$ & \multicolumn{2}{c}{ } & $19$&$938(4)$ \\

  A$^{0}_{v}$X$^{-}_{vc'c'}$ & $\rightarrow$ & A$^0_{v}+{}$X$^-_{vc'c'}$ & \multicolumn{2}{c}{ } & $21$&$46(3)$ \\

\hline

    X$^-_{vcc'}$ & $\rightarrow$ & X$_{vc}+{}$e$_{c'}$ &
    $6$&$0000(5)$ & \multicolumn{2}{c}{unbound} \\

    X$^-_{vcc}$ & $\rightarrow$ & X$_{v}+{}$e$_{c}$ &
    $9$&$1170(5)$ & $11$&$83(1)$\\

    D$^0_{c'}$X$_{vc}$ & $\rightarrow$ & X$_{vc}+{}$D$^0_{c'}$
    & $1$&$926(5)$ & \multicolumn{2}{c}{unbound}\\

    D$^-_{c'c'}$X$_{vc}$ & $\rightarrow$ & X$_{vc}+{}$D$^-_{c'c'}$ & $10$&$26(1)$ & \multicolumn{2}{c}{unbound}\\

    D$^-_{cc'}$ & $\rightarrow$ & D$^0_{c'}+{}$e$_{c}$ & $2$&$6020(5)$
    & \multicolumn{2}{c}{unbound}\\

    D$^0_{c'}$X$^-_{vcc}$ & $\rightarrow$ & D$^0_{c'}+{}$X$^-_{vcc}$ & $8$&$03(1)$ & \multicolumn{2}{c}{unbound}\\

    D$^-_{cc}$ & $\rightarrow$ & D$^0_{c}+{}$e$_{c}$ &
    $4$&$4297(9)$ & \multicolumn{2}{c}{unbound}\\

    D$^+$X$_{vc}$ & $\rightarrow$ & D$^++{}$X$_{vc}$ & $19$&$156(2)$
    & $1$&$286(8)$ \\

    \hline \hline
  \end{tabular}
\end{table}

It is clear from Table \ref{tab:dmc_dissociations} that the
approximate Keldysh interaction performs well at calculating binding
energies provided the dissociation does not involve significant
changes to short-range pair distributions.  As an extreme case, the
binding energy of an exciton, which is simply equal to its total
energy and hence does not benefit from any cancellation of errors, is
overestimated by 23\% when the approximate Keldysh interaction is
used.

  We are not aware of any published experimental
  results on donor-bound interlayer complexes in heterobilayers, but we discuss
  the validity of our results for intralayer complexes in Appendix
  \ref{sec:intralayer_comparison}.

\subsection{Sensitivity to model parameters} \label{sec:param_sens}

  We have performed test calculations to
  determine the sensitivity of the DMC-evaluated D$^0_{c'}$X$_{vc'}$
  binding energy to the model parameters $m_{c'}$, $m_v$, $r_*$,
  $r'_*$, and $d$ in the hBN-encapsulated heterobilayer. Note that
  $r_*$ and $r'_*$ are here the screening-length parameters in vacuum,
  so that the screening lengths in a dielectric environment are
  $r_*/\epsilon$ and $r'_*/\epsilon$.  We find that, upon variation of
  each of the parameters in turn by $\pm10\%$ from the values listed
  in Table \ref{tab:mat_parameters}, the D$^0_{c'}$X$_{vc'}$ binding
  energy never varies by more than $8\%$ (1.8 meV), as shown in Table
  \ref{tab:param_sens}.  The derivatives of the binding energy with
  respect to the parameters were evaluated numerically by the central
  difference approximation.  Nondimensionalizing lengths by the
  exciton Bohr radius and energies by the exciton
  Rydberg,\cite{mostaani_excitonic_prb_2017} it is easy to show that
  the derivative of a binding energy ${\cal E}^{\rm b}$ with respect
  to the dielectric constant is
\begin{equation} \frac{\partial \mathcal{E}^{\rm b}}{\partial
\epsilon} = -\frac{1}{\epsilon}\left(2r_*\frac{\partial
    \mathcal{E}^{\rm b}}{\partial r_*} + 2r'_*\frac{\partial
    \mathcal{E}^{\rm b}}{\partial r'_*}+d\frac{\partial
    \mathcal{E}^{\rm b}}{\partial d} +2\mathcal{E}^{\rm
    b}\right), \end{equation} allowing us to evaluate the sensitivity
of the binding energy with respect to the dielectric constant.  We
find that the binding energies are most sensitive to the screening
parameter $r_*$, followed by the dielectric constant $\epsilon$,
followed by the electron and hole masses $m_{c'}$ and $m_v$, and that
the sensitivity to the layer separation $d$ is relatively weak. The
sensitivity to the screening parameter $r'_*$ is very weak in the
present case, because only one hole resides in the WSe$_2$ layer.

\begin{table}[!]
  \centering
\caption{Derivatives of DMC binding energies $\mathcal{E}_{{\rm
      D}^0_{c'}{\rm X}_{vc'}}^{\rm b}$ of the interlayer donor-bound
  trion under variations $\delta P$ of different model parameters $P$.
  With the exception of the dielectric constant $\epsilon$, the
  parameters are varied by $\pm 10$\% about the values listed in Table
  \ref{tab:mat_parameters} and the central difference approximation is
  used to estimate the derivative with respect to the parameter value.
  The binding energy when all the parameters take the values listed in
  Table \ref{tab:mat_parameters} is $\mathcal{E}_{{\rm D}^0_{c'}{\rm
      X}_{vc'}}^{\rm b}=22.516(4)$ meV\@.  Note that $r_*$ and $r'_*$
  are here the screening lengths for a monolayer in vacuum.  The derivative
  of the binding energy with respect to the dielectric constant
  $\epsilon$ was evaluated by the chain rule, as described in the text.}
\label{tab:param_sens}
  \begin{tabular}{crcc}
  \hline \hline
 ~~$P$~~&~~$\delta P$~~~& $\mathcal{E}_{{\rm D}^0_{c'}{\rm X}_{vc'}}^{\rm b}$ (meV) &
  $\partial \mathcal{E}_{{\rm D}^0_{c'}{\rm X}_{vc'}}^{\rm b}/\partial P$\\ \hline
  \multirow{2}{*}{~~$m_{c'}$}&$+10\%$ & $23.27(1)$ & \multirow{2}{*}{$20.7$ meV/$m_0$} \\
   &$-10\%$& $21.70(1)$ & \\ \hline
  \multirow{2}{*}{~~$m_v$}&$+10\%$ & $22.71(1)$ & \multirow{2}{*}{$6.20$ meV/$m_0$} \\
   &$-10\%$& $22.29(1)$ & \\ \hline
  \multirow{2}{*}{~~$r_*$}&$+10\%$ & $20.96(1)$ & \multirow{2}{*}{$-0.421$ meV/\AA} \\
   &$-10\%$& $24.31(1)$ & \\ \hline
  \multirow{2}{*}{~~$r'_*$}&$+10\%$ & $22.46(1)$ & \multirow{2}{*}{$-0.00691$ meV/\AA} \\
   &$-10\%$& $22.52(1)$ & \\ \hline
  \multirow{2}{*}{~~$d$}&$+10\%$ & $22.93(1)$ & \multirow{2}{*}{$0.705$ meV/\AA}\\
   &$-10\%$& $22.02(1)$ & \\ \hline
  \multirow{2}{*}{~~$\epsilon$}&$+10\%$ & $\sim 20.97$ & \multirow{2}{*}{$-3.87$ meV}\\
   &$-10\%$& $\sim 24.06$ & \\ \hline \hline
  \end{tabular}
\end{table}

   We have also performed DMC calculations with
  $\epsilon = 4.5$ (instead of $\epsilon=4$), finding that the
  X$_{vc'}$, X$^-_{vc'c'}$, D$^0_{c'}$, and D$^0_{c'}$X$_{vc'}$
  binding energies are 76, 3.8, 207, and 20.7 meV, respectively.  This
  directly confirms that the sensitivity to the precise value of the
  dielectric constant of the environment is relatively weak.  The
  value of $\partial \mathcal{E}_{{\rm D}^0_{c'}{\rm X}_{vc'}}^{\rm
    b}/\partial \epsilon$ found by the forward difference
  approximation is $-3.63$ meV, which is in reasonable agreement with
  the value obtained using the chain rule, reported in Table
  \ref{tab:param_sens}.

\subsection{Comparison of intralayer binding energies with experimental results} \label{sec:intralayer_comparison}

For TMD monolayers, experimental agreement
  with QMC calculations of the binding energies of charge-carrier
  complexes employing the Keldysh interaction has previously
  been addressed in
  Refs.\ \onlinecite{marcin_bindingenergies_prb_2017} and
  \onlinecite{mostaani_excitonic_prb_2017}.  Trion binding energies
  are found to be in excellent agreement with experimental results.

Relatively few experimental studies of charge-carrier complexes in
heterobilayers have been performed to date. Ceballos \textit{et
  al.}\ studied a SiO$_2$/MoSe$_2$/MoS$_2$/vacuum
sample,\cite{ceballos2014ultrafast} performing PL measurements on
monolayer MoSe$_2$, monolayer MoS$_2$, and heterobilayer
MoSe$_2$/MoS$_2$ regions of their sample. Gong \textit{et
  al.}\ studied a SiO$_2$/MoS$_2$/WS$_2$/vacuum
sample,\cite{gong2014vertical} again performing PL measurements on
each of the three distinct surface regions. Both experiments, although
studying different TMD bilayers prepared by different means, observed
only small shifts in the dominant intralayer exciton lines on moving
from monolayer regions to bilayer regions. Our heterobilayer results
of Table \ref{tab:dmc_dissociations}, in conjunction with monolayer
binding-energy fitting formulas presented in
Ref.\ \onlinecite{mostaani_excitonic_prb_2017} provide further support
for this claim.  The intralayer exciton energy reported in Table
\ref{tab:dmc_total_energy} for an exciton X$_{vc}$ in the WSe$_2$
layer of a hBN/MoSe$_2$/WSe$_2$/hBN heterostructure is $-140.4$ meV,
whereas the exciton total energy in monolayer WSe$_2$ encapsulated in
hBN is $-159.7$ meV, according to the monolayer fitting formula.  The intralayer
negative trion X$^{-}_{vcc}$ binding energy reported in Table
\ref{tab:dmc_dissociations} is $11.8$ meV, whereas the
fitted negative-trion binding energy in monolayer WSe$_2$ encapsulated in hBN is
$13.6$ meV\@.  Thus the intralayer exciton energy differs by about
$19$ meV from the monolayer exciton energy, while the intralayer trion
binding energy differs by about 2 meV from the monolayer result.

In summary, intralayer binding energies in a heterobilayer are very
similar to monolayer binding energies, and hence the validity of our
model may be judged by examining previously reported results for TMD
monolayers.\cite{marcin_bindingenergies_prb_2017,mostaani_excitonic_prb_2017}

\subsection{Calculation of the overlap integrals}
\label{sub:overlaps}

\subsubsection{VMC evaluation of the normalization integral of a
many-body wave function} \label{sec:norm}

Consider a complex of $N$ quantum particles with unnormalized wave
function $\Phi({\bf R})$, where ${\bf R}=({\bf r}_1,\ldots,{\bf r}_N)$
is the $2N$-dimensional vector of all particle coordinates.  Let
$\Psi({\bf R})$ be a normalized, bound-state sampling wave function,
which ideally has a large overlap with $\Phi$ and the same asymptotic
behavior.  Then
\begin{eqnarray} \int |\Phi({\bf R})|^2 \, d^{2N}{\bf R} & = & \int |\Psi({\bf R})|^2 \left|\frac{\Phi({\bf R})}{\Psi({\bf R})}\right|^2 \, d^{2N}{\bf R} \nonumber \\ & = & \left< \left|\frac{\Phi({\bf R})}{\Psi({\bf R})}\right|^2 \right>_{|\Psi|^2}. \end{eqnarray}
Hence we can evaluate the normalization of $\Phi$ by VMC sampling of
$|\Psi({\bf R})|^2$.  We used the simple Jastrow form
\begin{equation} \Psi({\bf R})=\prod_{i=1}^N \left( \sqrt{\frac{2}{\pi}} c
e^{-cr_i} \right),
\end{equation}
for the sampling wave function, where the exponent $c$ is a positive,
adjustable parameter that was chosen to maximize the efficiency of the
calculation.

\subsubsection{Evaluation of overlap integrals} \label{sec:vmc_overlap}

Numerical estimates of the various overlap integrals in the
expressions for the radiative recombination rates of donor-bound
trions in a hBN/MoSe$_2$/WSe$_2$/hBN system are reported in Table
\ref{table:overlaps}.  The ground state $\chi_{\rm 1s}(\rr_{\text{e}})$ and the
first excited state $\chi_{\rm 2s}(\rr_{\text{e}})$ of the neutral
donor atom (D$^0_{c'}$) were calculated using a finite-element method
(see Appendix \ref{sec:finite}). Using a VMC-optimized trial wave
function $\Phi( \rr_{\text{h}}, \rr_{\text{e}_1}, \rr_{\text{e}_2})$
for the ground state of the donor-bound negative trion
(D$^0_{c'}$X$_{vc'}$), we employed a grid-based method to evaluate
those overlap integrals in Table \ref{table:overlaps} that can be
reduced to one-dimensional radial integrals.  The remaining integrals
were evaluated by a VMC method, as described below.

\begin{table}[tb]
  \caption{Overlap integrals required for calculations of radiative
    recombination rates. Calculations are performed for a
    hBN/MoSe$_2$/WSe$_2$/hBN system.  $\Phi(\rr_{\text{h}},
    \rr_{\text{e} 1}, \rr_{\text{e} 2})$ is the ground-state wave
    function of the donor-bound negative trion, with both donor and
    electrons in the MoSe$_2$ layer and the hole in the WSe$_2$ layer
    (D$^0_{c'}$X$_{vc'}$). $\chi_{\rm 1s}( \rr_{\text{e}})$ and
    $\chi_{\rm 2s}(\rr_{\text{e}})$ are the ground-state and
    first-excited-state (rotationally invariant) wave functions of the
    neutral donor atom in the MoSe$_2$ layer
    (D$^0_{c'}$). \label{table:overlaps}}
          {\renewcommand{\arraystretch}{2}
  \begin{tabular}{lcc}
    \hline \hline

    Overlap & Approx.\ Keldysh & Bilayer pot.\\

    \hline

    $\frac{| \Phi (\mathbf{0}, \mathbf{0}, \mathbf{0}) |^2}{\int | \Phi
      |^2 \, d^6 \mathbf{R}}$ & $1.29 \times 10^{- 9}$ {\AA}$^{-6}$ & $2.75
    \times 10^{- 9}$ {\AA}$^{- 6}$\\

    $\frac{\left| \int \Phi (\rr, \rr, \mathbf{0})
      \, d^2 \rr \right|^2}{\int | \Phi |^2 \, d^6 \mathbf{R}}$ &
    $8.09 \times 10^{- 3}$ {\AA}$^{- 2}$ & $6.08 \times 10^{- 3}$ {\AA}$^{- 2}$\\

    $\frac{\int | \Phi (\rr, \rr, 0) |^2 \, d^2
      \rr}{\int | \Phi |^2 \, d^6 \mathbf{R}}$ & $1.28 \times 10^{-
      6}$ {\AA}$^{- 4}$ & $1.38 \times 10^{- 6}$ {\AA}$^{- 4}$\\

    $\frac{\int | \Phi (\rr,\rr,\rr) |^2 \, d^2 \rr}{\int |\Phi|^2 \,
      d^6\mathbf{R}}$ & $3.22 \times 10^{-7}$ {\AA}$^{-4}$ & $2.37
    \times 10^{-7}$ {\AA}$^{-4}$ \\

    $\frac{\left| \int \chi_{\rm 1s} (\rr) \Phi (\mathbf{0},
      \mathbf{0},\rr) \, d^2 \rr \right|^2}{\int | \Phi |^2 \, d^6
      \mathbf{R} \times \int | \chi_{\rm 1s} |^2 \, d^2 \rr}$ & $6.94 \times
    10^{- 7}$ {\AA}$^{- 4}$ & $1.21 \times 10^{- 6}$ {\AA}$^{- 4}$\\

    $\frac{\left| \iint \chi_{\rm 1s} (\rr') \Phi (\rr,
      \rr, \rr') \, d^2 \rr \, d^2 \rr'
      \right|^2}{\int | \Phi |^2 \, d^6 \mathbf{R} \times \int | \chi_{\rm 1s} |^2
      \, d^2 \rr}$ & $3.54$ & $1.47$\\

    $\frac{ \int \left| \int \chi_{\rm 1s} (\rr') \Phi (\rr,
  \rr, \rr') \, d^2 \rr'
  \right|^2 d^2 \rr \,}{\int | \Phi |^2 \, d^6 \mathbf{R} \times \int | \chi_{\rm 1s} |^2
  \, d^2 \rr}$ & $5.90\times 10^{-4}\, {\Ams}^{-2}$ & $3.85\times 10^{-4}\ {\Ams}^{-2}$\\

    $\frac{\left| \int \chi_{\rm 2s} (\rr) \Phi (\mathbf{0},
  \mathbf{0},\rr) \, d^2 \rr \right|^2}{\int | \Phi |^2 \, d^6
  \mathbf{R} \times \int | \chi_{\rm 2s} |^2 \, d^2 \rr}$ & $2.01
\times 10^{- 8}$ {\AA}$^{- 4}$ & $1.13 \times 10^{- 7}$ {\AA}$^{- 4}$\\

$\frac{\left| \iint \chi_{\rm 2s} (\rr') \Phi (\rr,
  \rr, \rr') \, d^2 \rr \, d^2 \rr'
  \right|^2}{\int | \Phi |^2 \, d^6 \mathbf{R} \times \int | \chi_{\rm 2s} |^2
  \, d^2 \rr}$ & $0.0379$ & $0.0254$\\

    $\frac{ \int \left| \int \chi_{\rm 2s} (\rr') \Phi (\rr,
  \rr, \rr') \, d^2 \rr'
  \right|^2 d^2 \rr \,}{\int | \Phi |^2 \, d^6 \mathbf{R} \times \int | \chi_{\rm 2s} |^2
  \, d^2 \rr}$ & $1.04\times 10^{-5}\, {\Ams}^{-2}$ & $1.89\times 10^{-5}\ {\Ams}^{-2}$\\

    \hline \hline
  \end{tabular}}
\end{table}

Let $\Psi$ be a sampling wave function, as defined in Appendix
\ref{sec:norm}. The overlap of the trion wave function with the
donor-atom wave function when an electron and a hole are pinned
vertically above one another is
\begin{eqnarray} & & \iint \chi^*({\bf r}_1)
\Phi({\bf r}_2,{\bf r}_2,{\bf r}_1) \, d^2{\bf r}_1 \, d^2{\bf r}_2
\nonumber \\ & & {} = \int |\Psi({\bf R})|^2 \frac{\chi^*({\bf r}_2)
  \Phi({\bf r}_1,{\bf r}_2,{\bf r}_1)}{|\Psi({\bf R})|^2} \, d^4{\bf
  R} \nonumber \\ & & {} = \left< \frac{\chi^*({\bf r}_2)\Phi({\bf
    r}_1,{\bf r}_2,{\bf r}_1)}{|\Psi({\bf R})|^2}
\right>_{|\Psi|^2}. \label{eq:overlap_vmc} \end{eqnarray} The last
expression can readily be evaluated by VMC sampling of $|\Psi|^2$,
using accurate numerical representations of the donor-atom wave
function $\chi({\bf r}_{\rm e})$ obtained in the finite-element
calculations.

The overlap integrals are precise to at least three significant
figures; however there is an unknown error arising from the fact that
the trial wave function $\Phi(\rr_{\text{h}}, \rr_{\text{e}_1},
\rr_{\text{e}_2})$ only approximates the exact ground state.

\section{Radiative recombination assisted by short-range Coulomb interactions}\label{sec:amp}
Consider the wave function $\chi(\rr)$ for D${}_{c'}^0$ complexes in
the long-range (Keldysh) approximation described in
Sec.\ \ref{sec:asymptotics}. The complex state can be written in the
form of Eq.\ (\ref{eq:D0}), with the substitution $\tilde{\chi}_{\kk}
\longrightarrow \tilde{\chi}_{\kk}^0$, and short-range electrostatic
interactions and interlayer tunneling can be treated as perturbations
to this initial state. Setting $\tau'=\tau$ and $\sigma'=\sigma$ in
Eq.\ (\ref{eq:D0}), radiative decay is determined by the matrix
element $\langle \tau, \qq|\hat{H}_{\rm r}|\mathrm{D}^0;\kk_{\rm
  h}\rangle^{(2)}$, where $| \tau,\qq \rangle =
a_{\tau}^\dagger(\qq)\ket{\Omega}$ is the final state in which a
photon of momentum $\qq$ and the appropriate polarization $\tau$ has
been emitted after recombination of the bound electron with the
delocalized hole. The notation $\ket{A}^{(2)}$ indicates that the
state includes corrections up to second order in perturbation theory,
in this case from the interlayer tunneling ($\hat{H}_{\rm t}$) and
short-range interaction ($\hat{U}_{\text{intra}}^{>}$) terms.

\begin{figure}[b]
\begin{center}
\includegraphics[width=\columnwidth]{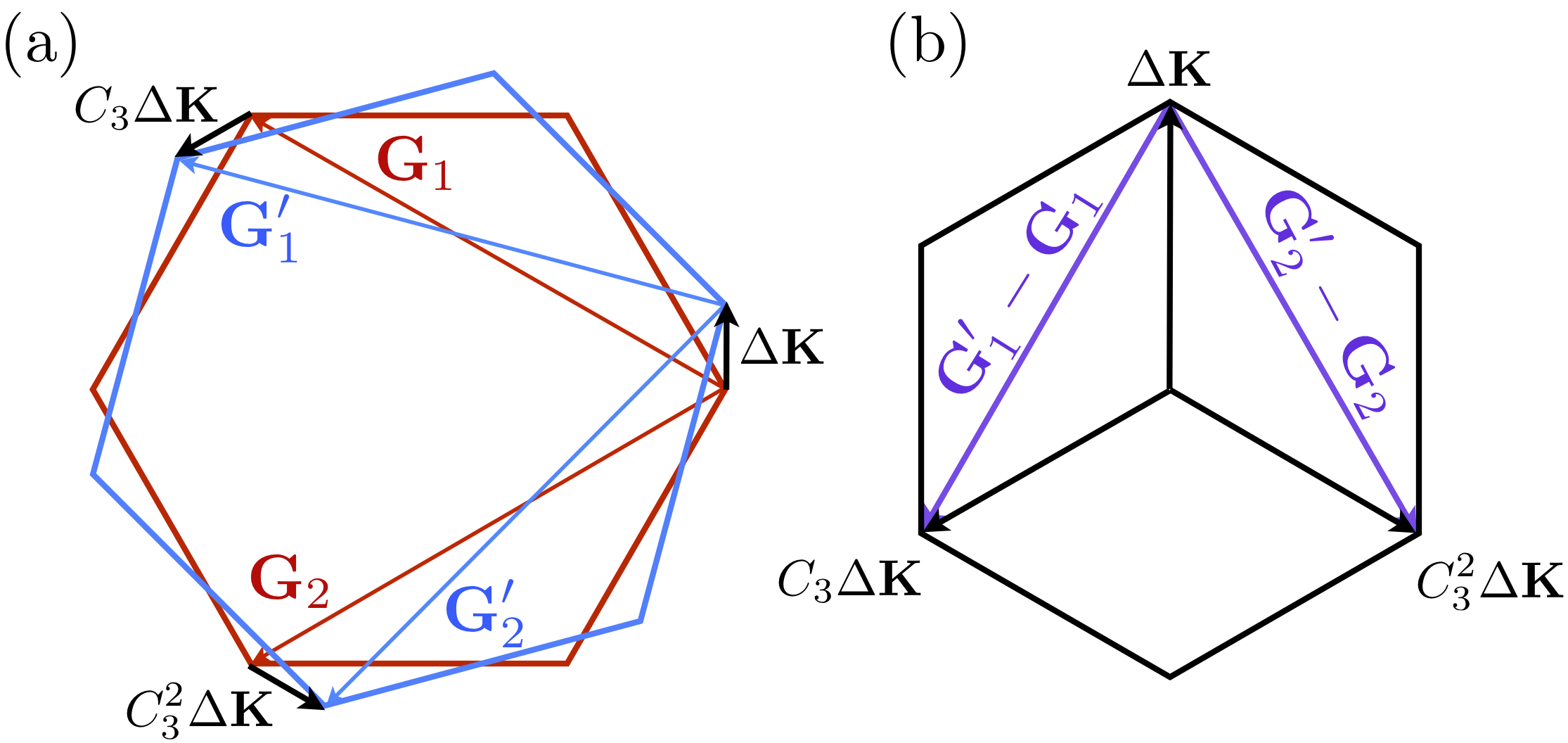}
\caption{(a) Reciprocal lattice vectors $\GG_n$ and $\GG_n'$ of the two layers, and valley mismatch vectors $C_3^n\Delta \KK$ ($n=0,\,1,\,2$), with the convention $\GG_0=\GG_0'=0$. (b) The vectors $C_3^n\Delta\KK$ are connected by $\GG_n'-\GG_n$.}
\label{fig:moire}
\end{center}
\end{figure}

The diagrams of Fig.\ \ref{fig:d0h_diagrams} correspond to those
corrections to the wave function that are relevant for radiative
recombination in the large-twist-angle regime, where ${\cal E}_{\rm
  D^0}^{\rm b}\ll \tfrac{\hbar^2\Delta K^2}{2m_\alpha}$. Following the
order of the diagrams in the figure, and assuming that $k_{\rm h},\,q
\ll \Delta K$, the optical matrix element for D$^0_{c'}$h$_{v}$
recombination is given in terms of the real-space impurity wave
function by
\begin{widetext}
\begin{equation}\label{eq:D0_omat}
{
\begin{split}
 & \langle \tau,\qq | \hat{H}_{\rm r} | \mathrm{D}^0;\kk_{\rm h} \rangle^{(2)}=\sum_{n=0}^2\frac{6\pi e^3\exp{-i\GG_n\cdot\rr_0}\chi(0)}{\epsilon\,r_*' S(C_3^n\Delta \KK)^2}\sqrt{\frac{4\pi \hbar c}{L\sqrt{q_\perp^2+q_\parallel^2}}}\\
 &\times\left[-\frac{\gamma t_{cc}/\hbar c}{\left({\cal E}_{\rm D^0}^{\rm b}+\Delta_c\right)\left({\cal E}_{\rm D^0}^{\rm b}+\tfrac{\hbar^2(C_3^n\Delta \KK)^2}{2m_{c'}}\right)} +\frac{\gamma't_{vv}/\hbar c}{\left({\cal E}_{\rm D^0}^{\rm b} +\Delta_v +\tfrac{\hbar^2(C_3^n\Delta \KK)^2}{2m_{c'}}+\tfrac{\hbar^2(C_3^n\Delta \KK)^2}{2m_{v'}} \right)\left({\cal E}_{\rm D^0}^{\rm b} +\tfrac{\hbar^2(C_3^n\Delta \KK)^2}{2m_{c'}}\right)} \right.\\
  &\qquad\quad- \left.\frac{\gamma't_{vv}/\hbar c}{\Delta_v \left(\Delta_v +
  \tfrac{\hbar^2(C_3^n\Delta \KK)^2}{2m_{v'}}\right)} + \frac{\gamma't_{vv}/\hbar
  c}{\left({\cal E}_{\rm D^0}^{\rm b} + \Delta_v +\tfrac{\hbar^2(C_3^n\Delta
  \KK)^2}{2m_{c'}}+\tfrac{\hbar^2(C_3^n\Delta \KK)^2}{2m_{v'}}\right)\left(\Delta_v +
  \tfrac{\hbar^2(C_3^n\Delta \KK)^2}{2m_{v'}}\right)}\right],
\end{split}
}
\end{equation}
\end{widetext}
where the Bragg vectors $\GG_n$ and valley mismatch momenta
$C_3^n\Delta\KK=\Delta\KK+(\GG_n'-\GG_n)$ are shown in Fig.\ \ref{fig:moire}.
The matrix element in Eq.\ \eqref{eq:D0_omat} can be written in terms of the stacking-dependent
function $F(\rr_0)=1+\exp{-i\GG_1\cdot\rr_0}+\exp{-i\GG_2\cdot\rr_0}$ (see Fig.\ \ref{fig:moire}). 
We additionally assume that the CB and VB spacings remain a large
scale in the problem, such that $\tfrac{\hbar^2\Delta K^2}{2m_\alpha}
\ll \Delta_c,\,\Delta_v$. In this approximation, the third and fourth terms in Eq.\ (\ref{eq:D0_omat})
cancel out, corresponding to the diagrams in Figs.\ \ref{fig:d0h_diagrams}(c)
and (d). Substituting the resulting expression into Eq.\ (\ref{eq:golden}) gives
Eq.\ (\ref{eq:d0h}), where the probability that the hole state is
occupied is introduced through the hole density $N(\kk_{\rm h})$. This
analysis can be carried out for D$^0_{c'}$X$_{vc'}$ complexes, yielding
Eqs.\ (\ref{eq:d0x1}) and (\ref{eq:d0x2}).

The large momentum components introduced by the short-range
interaction terms are irrelevant in the small-twist-angle regime,
which is dominated by the small momentum sector of the wave function. In this
case, the optical matrix element is obtained from the perturbed state
$| D^0;\kk_{\rm h} \rangle^{(1)}$, including first-order tunneling
corrections [Eq.\ (\ref{eq:first_order})]. The optical matrix element is
\begin{equation}
{
\begin{split}
  &\langle\tau,\qq|\hat{H}_{\rm r}|\mathrm{D}^0;\kk_{\rm
  h}\rangle=\sum_{n=0}^2\int d^2r\,\frac{\exp{i(C_3^n\Delta\KK+\kk_{\rm
  h}+\qq)\cdot\rr}}{\sqrt{S}} \chi(\rr)\\
  &\,\times\sqrt{\frac{4\pi\,\hbar c e^2}{SL\,q}}\left[\frac{t_{vv}\gamma'}{\hbar c\,\Delta_v} - \frac{t_{cc}\gamma}{\hbar c\left(\Delta_c+{\cal E}^{\rm b}_{\rm D^0}\right)} \right]\exp{-i\GG_n\cdot\rr_0}.
\end{split}
}
\end{equation}
Substituting into Eq.\ (\ref{eq:golden}) leads to
Eq.\ (\ref{eq:d0h_lesser}), and similar procedures are used to obtain
Eq.\ (\ref{eq:gd0xsmall}) for D$_{c'}^0$X$_{vc'}$ complexes. Notice that
the second radiative channel for D$_{c'}^0$X$_{vc'}$ discussed in the
main text does not apply to this regime. The small-twist-angle
analogue to the recoil process due to electron-electron interactions
involves a small momentum transfer, and is thus already contained in
the unperturbed state $|\mathrm{D}^0\mathrm{X}\rangle$.

\section{Phonon effects on radiative recombination}\label{app:phonon_assisted}

The discussion of Appendix \ref{sec:amp} can easily be adapted to
e-ph interactions, $\hat{H}_{\text{e-ph}}$ [Eq.\ (\ref{eq:e-p})]. In
the following we adopt the assumptions introduced in Appendix
\ref{sec:amp}; namely, $\tfrac{\hbar^2\Delta
  K^2}{2m_\alpha},\,{\cal E}_{\rm D^0}^{\rm b} \ll \Delta_c,\,\Delta_v$. In
addition, we use $\hbar\omega_{\Lambda,\nu}(\xi)\ll
\Delta_c,\,\Delta_v$, which is always valid in our cases of interest.

In the large-twist-angle regime, consider the process whereby the
electron in a D$_{c'}^0$ bridges the valley mismatch by emitting a
phonon in mode $\nu$ and momentum $\xxi$, with $\xxi \sim \Delta\KK$,
in either the electron ($\Lambda = {}$e) or hole ($\Lambda = {}$h)
layer. The electron recombines with a delocalized hole of momentum
$\kk_{\rm h}$, emitting a photon of momentum $\qq$ and polarization $\mu$,
leading to the final state
\begin{equation}
  \ket{\tau,\qq;\nu,\xxi}_{\Lambda} = a_{\tau}^\dagger(\qq)b_{\Lambda,\nu,-\xxi}^\dagger\ket{\Omega}.
\end{equation}
Considering the phonon energies presented in Table
\ref{tab:phonon_params}, in this regime we have $\tfrac{\hbar^2\Delta
  K^2}{2m_\alpha}\gg \hbar\omega_{\Lambda,\nu}$, and the radiative
matrix elements with phonon emission can be approximated by
\begin{subequations}
\begin{equation}\label{eq:matrix_d0h_phonons_gtr_h}
\begin{split}
  &{}_{\rm h}\langle \tau,\qq;\nu,\xxi |\hat{H}_{\rm r}|\mathrm{D}^0;\kk_{\rm h}\rangle \approx \,\frac{\tilde{\chi}(\kk_{\rm h}+\vec{\xxi}-C_3^n\Delta\KK)}{S}g_{v,\nu}(\Delta K)\\
  &\quad\times\sqrt{\frac{4\pi \hbar c}{SL\,q}}\frac{\gamma't_{vv}}{\hbar c\,\Delta_v}\left[\frac{2m_v}{\hbar^2\Delta K^2} - \frac{2m_{c'}}{\hbar^2\Delta K^2} \right]\exp{-i\GG_n\cdot\rr_0},
\end{split}
\end{equation}
\begin{equation}\label{eq:matrix_d0h_phonons_gtr_e}
\begin{split}
  &{}_{\rm e}\langle \tau,\qq;\nu,\xxi |\hat{H}_{\rm r}|\mathrm{D}^0;\kk_{\rm h}\rangle \approx \,\frac{\tilde{\chi}(\kk_{\rm h}+\vec{\xxi}-C_3^n\Delta\KK)}{S}g_{c',\nu}(\Delta K)\\
  &\quad\times\sqrt{\frac{4\pi \hbar c}{SL\,q}}\frac{\gamma t_{cc}}{\hbar c\,\Delta_c}\left[\frac{6m_{c'}}{\hbar^2\Delta K^2} - \frac{6m_v}{\hbar^2\Delta K^2} \right]\exp{-i\GG_n\cdot\rr_0}.
\end{split}
\end{equation}
\end{subequations}
For large twist angle, finite values of the wave function are obtained only if
$\xxi \approx C_3^n\Delta\KK$, resulting in three final phonon states distinguishable by the direction of their momenta, and
interference effects are lost. Furthermore, when substituting Eqs.\ (\ref{eq:matrix_d0h_phonons_gtr_h}) and
(\ref{eq:matrix_d0h_phonons_gtr_e}) into the golden rule [Eq.\ (\ref{eq:golden})], the
stacking-dependent phases also disappear.
The result is Eq.\ (\ref{eq:G_d0h_phonon_gtr}), and a similar procedure
leads to Eq.\ (\ref{eq:G_d0x_phonon_gtr}) for the phonon-assisted
decay of D$_{c'}^0$X$_{vc'}$ complexes.

The situation is more subtle in the small-twist-angle regime, where
interference effects are restored in processes within the hole layer, and
$\hbar\omega_{\Lambda,\nu} \lesssim \hbar^2\Delta
  K^2/(2m_\alpha)$, such that the phonon dispersion becomes important. The
optical matrix elements are
\begin{equation}
\begin{split}
  &{}_{\rm h}\langle \tau, \qq;\xxi,\nu|\hat{H}_{r}|\mathrm{D}^0;\kk_{\rm h}\rangle\approx\sum_{n=0}^2\frac{\tilde{\chi}(\kk+\xxi-C_3^n\Delta\KK)}{S}\\
  &\times\sqrt{\frac{4\pi \hbar c}{SL\,q}}\left[\frac{\gamma't_{vv}\,g_{v,\nu}(\xxi)}{\hbar c\,\Delta_v \left(\tfrac{\hbar^2\xi^2}{2m_v} + \hbar\omega_{\nu}(\xxi)\right)}\right.\\
  &\qquad\qquad-\left.\frac{\gamma t_{cc}\,g_{v,\nu}(\xxi)}{\hbar c\,\Delta_c\left(\tfrac{\hbar^2\xi^2}{2m_{v}} +\hbar\omega_{\nu}(\xxi)+ {\cal E}^{\rm b}_{\rm D^0} \right)}\right]\exp{-i\GG_n\cdot\rr_0},
\end{split}
\end{equation}
\begin{equation}
\begin{split}
  &{}_{\rm e}\langle\tau,\qq;\xxi,\nu|\hat{H}_{\rm r}|\mathrm{D}^0;\kk_{\rm h}\rangle=\sum_{n=0}^2\frac{\tilde{\chi}(\kk+\xxi-C_3^n\Delta\KK)}{S}\\
  &\times\sqrt{\frac{4\pi \hbar c}{SL\,q}}\left[\frac{\gamma t_{cc}g_{\nu,c'}(\xxi)}{\hbar c\,\Delta_c\left(\hbar\omega_{\nu}(\xxi)+{\cal E}^{\rm b}_{\rm D^0} \right)}\right.\\
  &\qquad\qquad\left.- \frac{\gamma't_{vv}g_{\nu,c'}(\xxi)}{\hbar c\,\Delta_v\left(\hbar\omega_{\nu}(\xxi)+{\cal E}^{\rm b}_{\rm D^0} \right)}\right]\exp{-i\GG_n\cdot\rr_0}.
\end{split}
\end{equation}
Following Ref.\ \onlinecite{wangyao_coupling}, we use $t_{cc}\ll t_{vv}$ to
simplify these expressions. Using Fermi's golden rule and integrating
over the photon momentum we obtain the decay rates
\begin{subequations}
\begin{equation}\label{eq:opt_d0h}
\begin{split}
  &\Gamma_{\rm D^0h}^{<,\nu}\approx\sum_{\vec{\xi}}\left[\frac{\abs{F(\rr_0)}^2\abs{g_{v,\nu}(0)}^2 }{\left(\tfrac{\hbar^2\xi^2}{2m_v} + \hbar\omega_{\nu}\right)^2}+\frac{3\abs{g_{c',\nu}(0)}^2}{\left(\hbar\omega_{\nu}+{\cal E}^{\rm b}_{\rm D^0} \right)^2}\right]\\
  &\times\int d^2r\int d^2r'\,\exp{i(\xxi-\Delta\KK)\cdot(\rr'-\rr)}\chi(\rr)\chi^{*}(\rr')\\
  &\times \frac{e^2}{\hbar c} \frac{4\tilde{E}_{\rm g}n_{\rm h}}{\hbar \,S}\left[\frac{\gamma't_{vv}}{\hbar c\,\Delta_v}\right]^2;\,\nu = \text{LO},\,\text{HP},
\end{split}
\end{equation}
\begin{equation}\label{eq:ac_d0h}
\begin{split}
  \Gamma_{\rm D^0h}^{<,\text{LA}}\approx&\,\sum_{\vec{\xxi}}\left[\frac{\abs{F(\rr_0)}^2\abs{g_{v,\text{LA}}(\xxi)}^2}{\left(\tfrac{\hbar^2\xi^2}{2m_v}+\hbar\,c_{\text{LA}}\xi\right)^2}+ \frac{3\abs{g_{c',\text{LA}}(\xxi)}^2}{{\cal E}^{\rm b}_{\rm D^0}{}^2 }\right]\\
  &\times\int d^2r\int d^2r'\,\exp{i\xxi\cdot(\rr'-\rr)}\chi(\rr)\chi^{*}(\rr')\\
  &\times \frac{e^2}{\hbar c}\frac{4\tilde{E}_{\rm g}n_{\rm h}}{\hbar \,S}\left[\frac{\gamma' t_{vv}}{\hbar c\, \Delta_v} \right]^2,
\end{split}
\end{equation}
\end{subequations}
for optical and acoustic phonon modes, respectively.

The divergence at $\xi=0$ in Eq.\ (\ref{eq:ac_d0h}) makes the first
term dominant in the sum over $\xxi$, and we can neglect the
second. The sum can be evaluated exactly in the continuous
limit. Defining $\mathcal{F}(x)=-x[Y_1(x)+H_{-1}(x)]$, where $H_{n}(x)$
and $Y_n(x)$ are the $n$th Struve function and Bessel function of the
second kind, respectively, we obtain
\begin{equation}
\begin{split}
  &\Gamma_{\rm D^0h}^{<,\text{LA}}\approx\,\frac{e^2}{\hbar c}\frac{\Xi_{v}^2\,\tilde{E}_{\rm g}m_vn_{\rm h}}{\hbar^3 \rho c_{\text{LA}}^2}\left[\frac{\gamma' t_{vv}}{\hbar c\, \Delta_v} \right]^2\abs{F(\rr_0)}^2\\
  &\times\int d^2r\int d^2r'\,\exp{i\xxi\cdot(\rr'-\rr)}\chi(\rr)\chi^{*}(\rr')\mathcal{F}\left(\tfrac{2m_v\hbar c_{\text{LA}}\abs{\rr'-\rr}}{\hbar^2}\right).
\end{split}
\end{equation}
From the values reported in Table \ref{tab:phonon_params} we find that
the function $\mathcal{F}$ in the integrand decays over a
characteristic length scale of $100$ nm, much greater than the spread
of the localized wave function $\chi(\rr)$. Therefore, to a good
approximation, we can substitute $\mathcal{F}(0) = 2/\pi$ to evaluate
the integral. The final results for all phonon modes considered in
Eqs.\ (\ref{eq:G_d0h_phonon_opt}) and (\ref{eq:G_d0h_phonon_ac}), and
Eqs.\ (\ref{eq:G_d0x_phonon_opt}) and (\ref{eq:G_d0x_phonon_ac}) are
obtained by a similar procedure.

\section{Finite-element calculation of two body states in heterobilayer system}
\label{sec:finite}
\begin{figure}[b]
  \centering
  \includegraphics[width=0.48\textwidth]{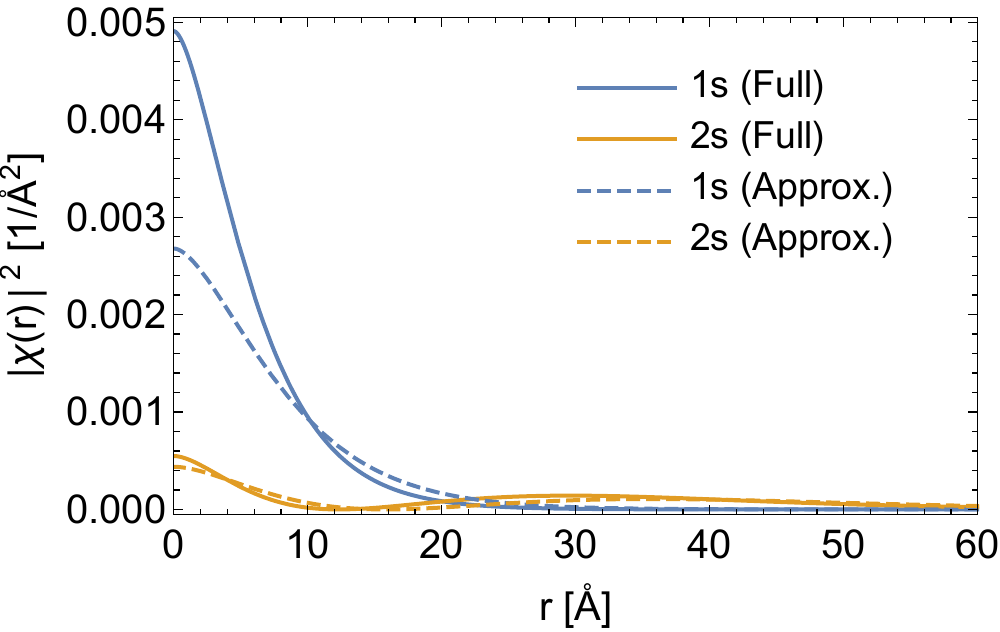}
  \caption{Probability distributions ($|\chi(r)|^2$) of the first two
    radially symmetric donor atom states in
    hBN/MoSe$_2$/WSe$_2$/hBN\@. The solid lines were obtained using
    the full bilayer potential Eq.\ (\ref{eq:exact_int_11}), and correspond to states with binding energies $\mathcal{E}^{\rm b}_{\rm 1s}=-229.03\,{\rm meV}$ and $\mathcal{E}^{\rm b}_{\rm 2s}=-61.73\,{\rm meV}$.
    The dashed lines were obtained using the approximate
    intralayer Keldysh form Eq.\ (\ref{eq:approx_intralayer}).}
  \label{fig:X}
\end{figure}
The Schr\"odinger equation for two particles interacting through a
radially symmetric potential $\mathcal{U}(r)$ is given by
\cite{marcin_bindingenergies_prb_2017},
\begin{equation}
\left[-\frac{\hbar^2}{2m_{\rm 1}}\nabla_{\rm e}^2-\frac{\hbar^2}{2m_{\rm 2}}\nabla_{\rm h}^2-e^2\mathcal{U}(r_{\rm 12})\right]\Psi = E\Psi,
\end{equation}
where the form of the interaction $\mathcal{U}$ between charge
carriers is explained in Appendix \ref{sub:bilayer_keldysh}, depending
on the layer in which each particle is found.

Transforming the coordinates to the relative motion ${\bf r}={\bf
  r}_{\rm 1}-{\bf r}_{\rm 2}$ and the center-of-mass motion ${\bf
  R}=\frac{m_{\rm 1}{\bf r}_{\rm 1}+m_{\rm 2}{\bf r}_{\rm 2}}{m_{\rm
    1}+m_{\rm 2}}$ allows separation of the Schr\"{o}dinger equation
to the center-of-mass part whose solution is given by the plane wave
$\phi(R)=\frac{1}{\sqrt{S}}e^{i{\bf K}\cdot {\bf R}}$ and the energy $E=\frac{\hbar^2
  K^2}{2(m_{\rm 1}+m_{\rm 2})}$, and the relative-motion part given by
\begin{equation}
\left[-\frac{\hbar^2}{2\mu}\nabla^2-e^2\mathcal{U}(r)\right]\Psi = E\Psi,
\end{equation}
where $\mu=m_{\rm 1}m_{\rm 2}/(m_{\rm 1}+m_{\rm 2})$ is the reduced mass.

Transforming the equation into dimensionless quantities
\cite{marcin_bindingenergies_prb_2017, mostaani_excitonic_prb_2017}
using the excitonic Bohr radius $a_0^* = \frac{\epsilon \hbar^2}{\mu
  e^2}$ and the excitonic Rydberg energy $R_{\rm y}^*=\frac{\mu
  e^4}{2\epsilon^2 \hbar^2}$ gives
\begin{equation}
\left[-\tilde{\nabla}^2
  -\frac{1}{R_{\rm y}^*}\mathcal{U}(a_0^*\tilde{r})\right]\Psi
= \tilde{E}\Psi.
\label{eq:red}
\end{equation}
where $\tilde{r}=r/a_0^*$ and $\tilde{E}=E/R_{\rm y}^*$.  Using
separation of variables the general solution is given by
\begin{equation}
\Psi({\bf r}) = R(r)\Phi(\phi),
\end{equation}
where the angular-part solution is
\begin{equation}
\Phi(\phi) =\frac{1}{\sqrt{2\pi}}e^{i l \phi}.
\end{equation}
$l=0,\pm1,\pm2,\dots$ is the azimuthal quantum number with $\Phi(\phi)$
being an eigenfunction of the angular momentum operator
$L_z=-i\hbar\frac{\partial}{\partial \phi}$ with eigenvalue $\hbar l$.
The equation for the radial part  is
\begin{equation}
-R''(\tilde{r})-\frac{1}{\tilde{r}}R'(\tilde{r})+\frac{l^2}{\tilde{r}^2}R(\tilde{r})-\tilde{v}(\tilde{r})
R(\tilde{r}) = \tilde{E} R(\tilde{r}), \label{eq:rad_part_ex}
\end{equation}
where $\tilde{v}(\tilde{r})=\mathcal{U}(a_0^* \tilde{r})/R_{\rm y}^*$.
To solve Eq.\ (\ref{eq:rad_part_ex}) numerically we use the
substitution $u(\tilde{r})=R({\tilde r})\tilde{r}$, allowing us to impose
Dirichlet boundary conditions: $u(\tilde{r})=0$ at $\tilde{r}=0$ and
$\tilde{r}=\infty$. The equation can be solved using the
finite-element method implemented in Mathematica \cite{Mathematica}.
For the charged donor interacting with an electron in the ${\rm
  MoSe_2}$ layer, we have $\mu=m_c'$, and we solve
Eq.\ (\ref{eq:rad_part_ex}) using both the approximate Keldysh
interaction and the full bilayer potential for the intralayer
interaction between the donor and electron.  The normalized
probability distributions for the first two radially symmetric states
($1s, 2s$) obtained using both potentials are plotted in
Fig.\ \ref{fig:X}.

\end{document}